\documentclass[draft]{agujournal2019}
\usepackage{url} 
\usepackage{lineno}
\usepackage[inline]{trackchanges} 
\usepackage{soul}
\usepackage{color}
\usepackage{epstopdf} 
\draftfalse

\journalname{JGR: Earth Surface}

\begin{document}

%
%

\title{Morphodynamics of barchan-barchan interactions investigated at the grain scale}

%
%

\textcolor{blue}{An edited version of this paper was published by AGU. Copyright (2021) American Geophysical Union. Assis, W. R. and Franklin, E. M. (2021). Morphodynamics of barchan-barchan interactions investigated at the grain scale. Journal of Geophysical Research: Earth Surface, 126, e2021JF006237. https://doi.org/10.1029/2021JF006237.\\
To view the published open abstract, go to https://doi.org/10.1029/2021JF006237.}




\authors{W. R. Assis\affil{1}, E. M. Franklin\affil{1}}


\affiliation{1}{School of Mechanical Engineering, UNICAMP - University of Campinas,\\
Rua Mendeleyev, 200, Campinas, SP, Brazil}




\correspondingauthor{Erick M. Franklin}{erick.franklin@unicamp.br}




\begin{keypoints}
\item We determine the trajectories of individual grains during barchan-barchan interactions
\item We show the origin and destination of moving grains and typical lengths and velocities
\item We find the spreading rate of grains over the target barchan once dune-dune collision has occurred
\end{keypoints}

%
%

%
%


\begin{abstract}
Corridors of size-selected crescent-shaped dunes, known as barchans, are commonly found in water, air, and other planetary environments. The growth of barchans results from the interplay between a fluid flow and a granular bed, but their size regulation involves intricate exchanges between different barchans within a field. One size-regulating mechanism is the binary interaction between nearby dunes, when two dunes exchange mass via the near flow field or by direct contact (collision). In a recent Letter \cite{Assis}, we identified five different patterns arising from binary interactions of subaqueous barchans, and proposed classification maps. In this paper, we further inquire into binary exchanges by investigating the motion of individual grains while barchans interact with each other. The experiments were conducted in a water channel where the evolution of pairs of barchans in both aligned and off-centered configurations was recorded by conventional and high-speed cameras. Based on image processing, we obtained the morphology of dunes and motion of grains for all interaction patterns. We present the trajectories of individual grains, from which we show the origin and destination of moving grains, and their typical lengths and velocities. We also show that grains from the impacting dune spread with a diffusion-like component over the target barchan, and we propose a diffusion length. Our results provide new insights into the size-regulating mechanisms of barchans and barchanoid forms found on Earth and other planets.
\end{abstract}

\section*{Plain Language Summary}
Barchans are dunes of crescentic shape that are commonly found on Earth, Mars and other celestial bodies. Although of similar shape, their scales vary with the environment they are in, going from the millennium and kilometer for Martian barchans, down to the minute and centimeter in the aquatic case, passing by hundreds of meters and years for aeolian barchans. Other common characteristic is that barchans are organized in dune fields, where barchan-barchan collisions are an important mechanism for their size regulation. We took advantage of the smaller and faster scales of subaqueous dunes and performed experiments in a water channel, which allowed us to determine the trajectories of individual grains while two barchans interacted with each other, something unfeasible from field measurements on terrestrial or Martian deserts. We show typical lengths and velocities of individual grains, and that, in case of barchan collisions, grains from the impacting barchan spread with a diffusive component over the other barchan. Our results provide new insights into the evolution of barchans found in water, air, and other planetary environments.

\section{Introduction}

Fields of barchan dunes, crescent-shaped dunes with horns pointing downstream, are commonly found in different environments, such as rivers, Earth's deserts and on the surface of Mars \cite{Bagnold_1, Herrmann_Sauermann, Hersen_3, Elbelrhiti, Claudin_Andreotti, Parteli2}, being characterized by corridors of size-selected barchans. The growth of dunes results from the interplay between a fluid flow and a granular bed, with sand being transported as a moving layer called bedload. Barchan dunes usually appear under a one-directional fluid flow and limited sand supply \cite{Bagnold_1}, but the regulation of their size involves intricate interactions between different barchans within a field \cite{Hersen_2, Hersen_5, Kocurek, Genois, Genois2}. Barchan fields observed in nature result thus from complex interactions between a fluid flow, a sand bed, and existing bedforms.

\citeA{Hersen_2} showed that an isolated barchan within a dune field is marginally stable, since it receives and loses sand in proportion to its width and size of horns, respectively, meaning that the net flux of sand is positive for large barchans and negative for small ones. In addition, because smaller dunes move faster than larger ones \cite{Bagnold_1}, collisions could lead to a coarsening of the barchan field. \citeA{Elbelrhiti} showed that, in fact, barchan-barchan collisions and changes in wind direction induce surface waves that propagate faster than the barchan itself, which can regulate the size of barchans. If the barchan dune is larger than the characteristic length of the surface waves, the latter propagate toward one of the barchan horns and new barchans are ejected, a mechanism known as calving. Otherwise, calving is not observed (in case of a barchan-barchan collision, the two barchans simply merge). Later, \citeA{Worman} proposed that the wake of an upstream barchan can lead to calving on a downstream dune prior (or even without) a barchan-barchan collision, due to the same wave mechanism shown by \citeA{Elbelrhiti}.

The first studies on barchan interactions were based on field measurements of aeolian barchans, such as done by \citeA{Norris} and \citeA{Gay}. Field measurements are still important in investigating barchan interactions \cite{Vermeesch, Elbelrhiti2, Hugenholtz}, having shown that size regulation and the appearance of barchanoid forms are highly influenced by barchan-barchan collisions. However, given the long timescales in the aeolian case (of the order of the decade), time series for barchan collisions in aeolian fields are frequently incomplete, and conclusive results would need around a century to be achieved. In order to overcome this problem, numerical and experimental investigations were carried out over the last decades.

The numerical investigations were conducted using simplified models, both continuum \cite{Schwammle2, Duran2, Zhou2} and discrete \cite{Katsuki}, and most of them incorporated a few rules of barchan interactions in order to inquire into the mechanisms of sand distribution and evolution of dune fields. In particular, \citeA{Lima} and \citeA{Partelli5} proposed a simple model based on the inter-dune sand flux and a rule for the merging (coalescence) of dunes, \citeA{Lima} investigating barchan dunes in two dimensions and \citeA{Partelli5} transverse dunes in one dimension. As results, \citeA{Lima} showed that barchans reach eventually comparable sizes and are confined to corridors, and \citeA{Partelli5} that transverse dunes reach both the same heights and velocities. Later, \citeA{Katsuki2} and \citeA{Duran3} carried out numerical simulations to investigate the outcome of barchan-barchan collisions, from which they obtained the merging and exchange patterns (described in what follows). In addition, \citeA{Duran3} proposed an equation for the size distribution of barchans based on a balance of sand flux and a collision model. In the same line of \citeA{Duran3}, \citeA{Genois2} proposed an agent-based model using the balance of sand fluxes and elementary rules for barchan-barchan collisions that included a fragmentation-exchange pattern (described next) in addition to the merging and exchange ones. As general results, the models of \citeA{Duran3} and \citeA{Genois2} found that sand distribution due to collisions is a mechanism that explains the existence of corridors of size-selected barchans, with sparse and large or dense and small barchans. Different from previous works, \citeA{Bo} simulated numerically the growth and evolution of a barchan field using a scale-coupled model \cite{Zheng} in order to obtain the probability of barchan-barchan collisions. They found the probabilities for the occurrence of three collision patterns (merging, exchange and fragmentation-exchange, described next), and showed that probabilities vary with the flow strength, grain diameter, grain supply and height ratio of barchans. However, although varying several parameters, the authors did not investigate the mechanics of collisions, which remains to be fully understood.

In common, previous numerical works pointed toward homogeneous fields, but, although those investigations reproduced some collision types, model simplifications prevented them from reproducing correctly all existing short-range interactions (including collisions). Being more specific, the interactions strongly influenced by wake effects (chasing and fragmentation-chasing, described next), for which collision does not occur, are not explicitly dealt with, and the effects of grain types and flow conditions are not taken into account. Besides, numerical studies at the grain scale, showing the trajectories of individual grains, do not exist at the moment. Some of these aspects were only investigated recently (experimentally) at the bedform scale \cite{Assis}, and there is a complete lack of information at the grain scale.

Given the relatively fast and small scales of the subaqueous case (in the order of minutes and centimeters), the experiments on barchan-barchan interactions were conducted in water tanks and channels (with the exception of \citeA{Palmer}). Part of them investigated the disturbances in the fluid flow as two barchans approach each other, which may affect greatly bedload and surface erosion. In a sequence of experimental works, \citeA{Palmer} investigated the flow disturbances caused by an upstream barchan upon a downstream one in a wind tunnel when they are in an aligned configuration, for which they varied the volume ratios and fixed the longitudinal separation, and \citeA{Bristow}, \citeA{Bristow2} and \citeA{Bristow3} investigated the off-centered configuration in a water channel, where they fixed the volume ratio and varied the longitudinal separations. The experiments made use of particle image velocimetry (PIV), and found that the wake of the upstream dune increases turbulence levels on the downstream stoss surface, causing thus a larger erosion on the downstream dune, and that the transverse offset creates a channeling effect around one of the horns of the downstream barchan, promoting dune asymmetry. They showed also that near-bed fluctuations are particularly increased at the reattachment point and that streamwise vortices emerge from the horns, which can enhance even more erosion on the downstream barchan depending on the relative positions of dunes.

Another part of experiments were concerned with the bedform evolution as two dunes interacted with each other. In particular, \citeA{Endo2} and \citeA{Hersen_5} investigated barchan-barchan collisions, the former using a water flume to study the collisions of aligned barchans and the latter a tank in which the motion of a tray created a relative flow between the water and the bedform to investigate the collisions of off-centered barchans. While \citeA{Endo2} varied the mass ratio of barchans and kept the water flow rate, initial conditions and grain types fixed, \citeA{Hersen_5} varied the transverse distance of colliding dunes (referred to as impact or offset parameter) and maintained the other parameters fixed. In this way, these works complemented each other to a certain extent and showed, as main results, that barchan-barchan collisions produce smaller dunes, promoting sand redistribution. In addition, \citeA{Endo2} identified three types of collision patterns, which they named absorption, ejection and split, and which we call merging, exchange and fragmentation-chasing \cite{Assis} and explain in what follows. For other kinds of dunes, \citeA{Bacik} investigated the interaction between a pair of two-dimensional dunes in a narrow Couette-type circular channel, where, under the action of a turbulent flow, the pair of dunes interacted with each other over long times. Under such spanwise confinement, they found that turbulent structures of the flow induce a dune-dune repulsion that prevents dune collisions. They conjectured that such mechanism could happen for the interaction of two barchans of comparable size.

In spite of all those findings, a general picture for all barchan-barchan interactions was still lacking, i.e., the identification and organization of all interaction patterns in a parameter space including all relevant parameters: initial separation and alignment, dune masses, grain properties, and fluid velocity. In a recent paper \cite{Assis}, we investigated experimentally the short-range binary interactions of subaqueous barchans, including collisions, in both aligned and off-centered configurations. The experiments were conducted in a transparent channel where controlled grains were entrained by the water flow, forming a pair of barchans that interacted with each other. We varied the water flow rates, grain types (diameter, density and roundness), pile masses, longitudinal and transverse distances, and initial conditions. As a result, we identified five interaction patterns for both aligned and off-centered configurations and proposed two maps that provide a comprehensive classification for barchan-barchan interactions based on the ratio between the number of grains of each dune, Shields number and alignment of barchans. The five different patterns observed were classified as (i) chasing, when the upstream barchan does not reach the downstream one; (ii) merging, when the upstream barchan reaches the downstream one and they merge; (iii) exchange, when, once the upstream barchan reaches the downstream one, a small barchan is ejected; (iv) fragmentation-chasing, when the downstream dune splits before being reached by the upstream barchan and the new dunes outrun the upstream one; and (v) fragmentation-exchange, when fragmentation initiates, the upstream barchan reaches the splitting dune, and, once they touch, a small barchan is ejected. In addition, we showed that an ejected barchan has roughly the same mass of the impacting one and that the asymmetry of the downstream barchan is larger in wake-dominated processes. However, details at the grain scale of barchan-barchan interactions were not investigated.

Although previous studies have shown that barchan-barchan collision can be a size-regulating mechanism and identified the interaction patterns, none of them investigated the mass transfers between barchans prior and during collisions, the motion of grains once collision took place, nor, with the exception, partially, of \citeA{Assis}, the dune morphodynamics during collisions. Therefore, mass transfers and motions at the grain scale during barchan-barchan interactions remain completely unknown. In this paper, we further inquire into barchan-barchan interactions by investigating the motion of grains while barchans interact with each other, allowing us to compute the mass exchanged between barchans and lost by the system, and the spreading of grains of an impacting barchan over the target one. The experiments were conducted in a water channel where the evolution of pairs of barchans in both aligned and off-centered configurations was recorded by conventional and high-speed cameras. Based on image processing, we tracked bedforms and grains for all interaction patterns. We present the trajectories of individual grains during different stages of barchan-barchan interactions, from which we find the origin and destination of moving grains, their typical lengths and velocities, and the proportions of grains exchanged between barchans and lost by the entire system. We also show that grains from the impacting dune spread with a diffusion-like component over the target barchan, and propose a diffusion length for their dispersion. The present results provide new insights into the shape and size variations of barchans and barchanoid forms found in water, air, and other planetary environments.

In the following, Sec. \ref{sec:Exp} describes the experimental setup and procedure, Sec. \ref{sec:Res} presents the obtained results, and Sec. \ref{sec:Conclu} presents the conclusions.

\section{\label{sec:Exp} Experimental Setup}

\begin{figure}
\centering
\includegraphics[width=0.99\linewidth]{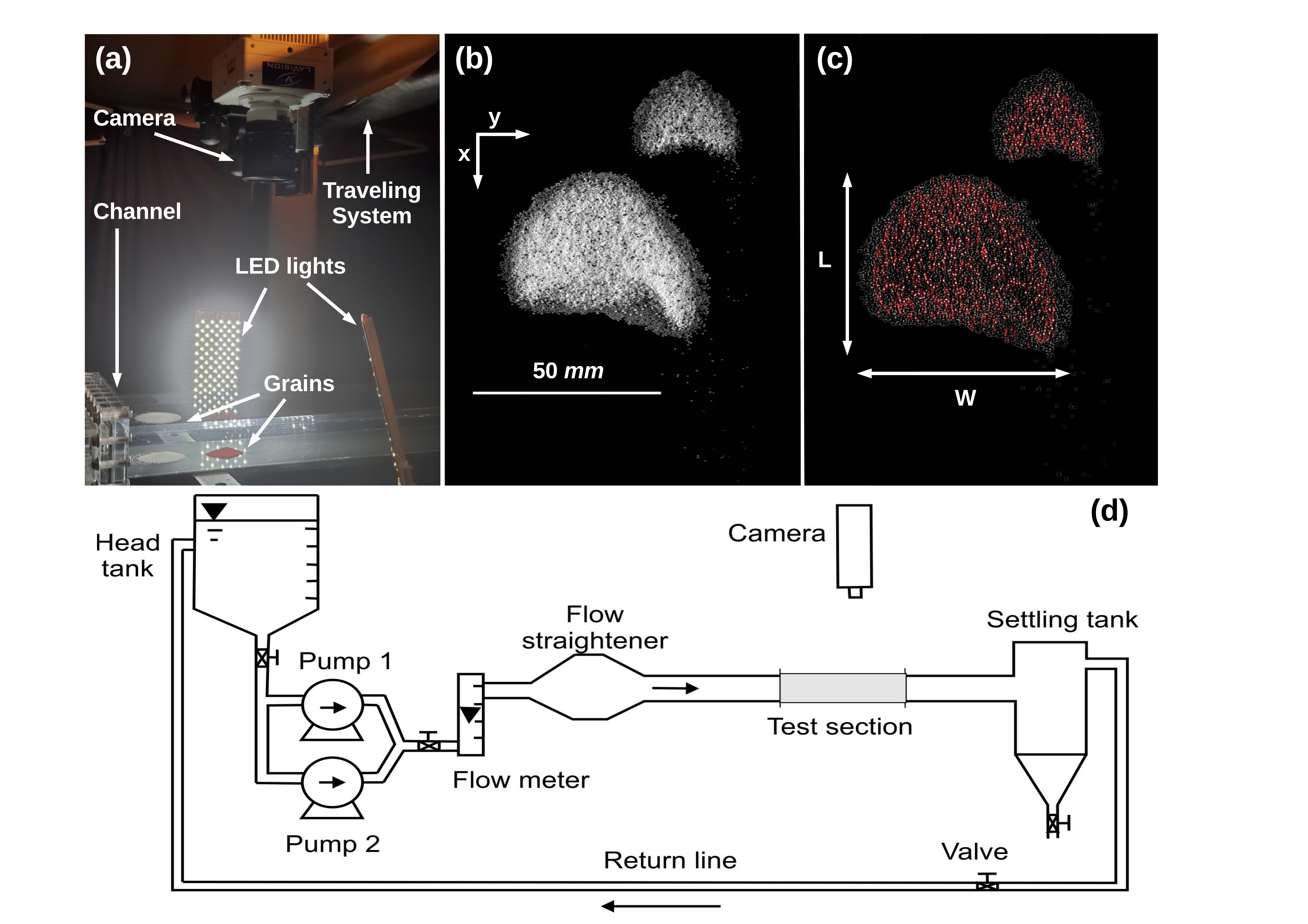}
\caption{Experimental setup, barchans and grains detection, and definition of some geometrical parameters. (a) Photograph of the experimental setup showing the test section, camera, traveling system, LED lights, and dunes on the bottom wall of the channel. (b) Top-view image of two interacting barchans, the water flow is from top to bottom. (c) Binarized image of interacting barchans showing identified grains that were tracked along images and some of the barchan dimensions. (d) Layout of the experimental setup.}
	\label{fig:setup}
\end{figure}

The experimental device is the same as in \citeA{Assis}, consisting of a water reservoir, two centrifugal pumps, a flow straightener, a 5-m-long closed-conduit channel, a settling tank, and a return line, where a pressure-driven water flow was imposed in the order just described. The channel was made of transparent material and had a rectangular cross section 160 mm wide and 2$\delta$ = 50 mm high, its 1-m-long test section starting 3 m downstream of the channel inlet. This corresponds to 40 hydraulic diameters, which assured a developed channel flow upstream the bedforms. The remaining 1-m-long section connected the exit of the test section to the settling tank. Figures \ref{fig:setup}d and \ref{fig:setup}a present, respectively, the layout of the experimental device and a photograph of the test section.

Controlled grains were poured inside the channel, filled previously with water, forming two conical piles that were afterward deformed into barchans by the imposed water flow. The pairs of bedforms were formed in either aligned or off-centered configurations and the longitudinal distance between initial piles was of the order of the diameter of the upstream pile. The size of the upstream dune (impact dune) was always equal or lesser than that of the downstream dune (target dune), since the dune velocity varies inversely with its size \cite{Bagnold_1}, their mass ratio varying within 0.021 and 1. We did not impose an influx of grains coming from regions upstream the impact dune, so that the entire system lost grains and decreased in mass along time. With that procedure, we obtained binary interactions for all five patterns described in \citeA{Assis}, in both aligned and off-centered configurations.

The ensemble of tests used tap water at temperatures within 25 and 28 $^{\circ}$C and round glass beads ($\rho_s$ = 2500 kg/m$^3$) with diameters $0.15$ mm $\leq\,d_s\,\leq$ $0.25$ mm and $0.40$ mm $\leq\,d_s\,\leq$ $0.60$ mm (not mixed with each other). In the following, we consider $d$ as the mean value of $d_s$. In order to facilitate the tracking of grains, tests focused on the mass exchange between barchans used 96-98 \% of white grains and 4-2 \% of black grains, for both dunes, and tests focused on particle diffusion at the grain scale used white grains for the impact and red grains for the target dune (colors inverted with respect to \citeA{Assis}), all of them with the same density, diameter and roundness for a given test. The cross-sectional mean velocity of water, $U$, was fixed at either 0.243 or 0.278 m/s (computed as the measured flow rate divided by the cross-sectional area), corresponding to Reynolds numbers based on the channel height, Re = $\rho U 2\delta /\mu$, of $1.22$ $\times$ $10^4$ and $1.39$ $\times$ $10^4 $, respectively, where $\mu$ is the dynamic viscosity and $\rho$ the density of the fluid. The shear velocities on the channel walls in the absence of dunes, $u_*$, were computed based on measurements with a two-dimensional two-component particle image velocimetry (2D2C-PIV) device and found to follow the Blasius correlation \cite{Schlichting_1}, being 0.0141 and 0.0159 m/s for the two imposed water flows. By considering the fluid velocities applied to each grain type, the Shields number, $\theta = (\rho u_*^2)/((\rho_s - \rho )gd)$, varied within 0.027 and 0.086, where $g$ is the acceleration of gravity. Because the shear velocity varies over the surface of each dune, as well as in some regions on the channel walls when in the presence of barchans \cite{Bristow, Bristow2, Bristow3}, we use $u_*$ (undisturbed by dunes) as the reference value for the fluid shearing. Microscopy images of the used grains and a table summarizing the tested conditions are available in the supporting information.

The evolution of bedforms was recorded by either a high-speed or a conventional camera mounted on a traveling system and placed above the channel, both the camera and traveling system being controlled by a computer. The high-speed camera was of complementary metal-oxide-semiconductor (CMOS) type with maximum resolution of 2560 px $\times$ 1600 px at 800 Hz, and we set its region of interest (ROI) within 2176 px $\times$ 960 px and 2560 px $\times$ 1600 px and the frequency to 200 Hz. The field of view varied from 117 mm $\times$ 75 mm to 205 mm $\times$ 112 mm, the area covered by each grain varying within 6 to 32 px in the images. The conventional camera, also of of CMOS type, had a maximum resolution of 1920 px $\times$ 1080 px at 60 Hz, which were the ROI and frequency set in the tests. For the tests on the exchange pattern, the field of view was 160 mm $\times$ 90 mm, the area covered by each grain ($d$ = 0.2 mm) corresponding thus to approximately 5 px, while the tests on the merging pattern had a field of view of 260 mm $\times$ 146 mm, the area covered by each grain ($d$ = 0.5 mm) corresponding to approximately 11 px. We mounted lenses of $60$ mm focal distance and F2.8 maximum aperture on the cameras and made use of lamps of light-emitting diode (LED) branched to a continuous-current source to provide the necessary light while preventing beating with the cameras. The conversion from px to a physical system of units was made by means of a scale placed in the channel previously filled with water. Movies showing the motion of grains over approaching and colliding barchans are available in the supporting information.

The acquired images were processed by numerical scripts written in the course of this work and based on \citeA{Crocker}, \citeA{Kelley}, \citeA{Houssais_1} and \citeA{Cunez3}. They basically removed the image background, binarized the images, identified the barchan morphology and individual grains, and computed the main morphological properties of bedforms, their relative distances and the motion of grains. Figures \ref{fig:setup}b and \ref{fig:setup}c present, respectively, raw and processed images, the latter showing identified grains that were tracked along images.

Given its high frequencies, the high-speed camera uses an internal memory to store the acquired images, to be discharged to a computer once the measurements are over or the memory full. Depending on the tests, the time for discharging image files was greater than that for reaching the next stage of interaction between dunes. These were the cases of tests with higher velocities ($U$ = 0.278 m/s), for which once the images were discharged we had to restart the tests from the beginning, under the same conditions, until reaching the next stage to be recorded. For the other tests, measurements were made in a continuous mode, the camera having discharged the files to the computer before the next stage was reached. We note that in spite of presenting the realization of one instance of each interaction stage, we recorded a large number of tests at normal (60 Hz) frequencies (123 of them presented in \citeA{Assis}) and repeated the data acquisition of all tests at higher frequencies. We verified that the trajectories were consistent with the results presented here and, because of the large amount of data presented in this paper (for instance, 22 movies in the supporting information), we did not fully process all the data and do not show all of them here.

\section{\label{sec:Res} Results and discussion}

Because in \citeA{Assis} the ensemble of tests for each pattern showed the same behavior, we present next the motion of grains for one instance of each pattern, in both aligned and off-centered configurations.

\subsection{\label{sec:Res_trajectories} Trajectories of grains leaving dunes and mass exchange}

\begin{figure}
\centering
\includegraphics[width=0.8\linewidth]{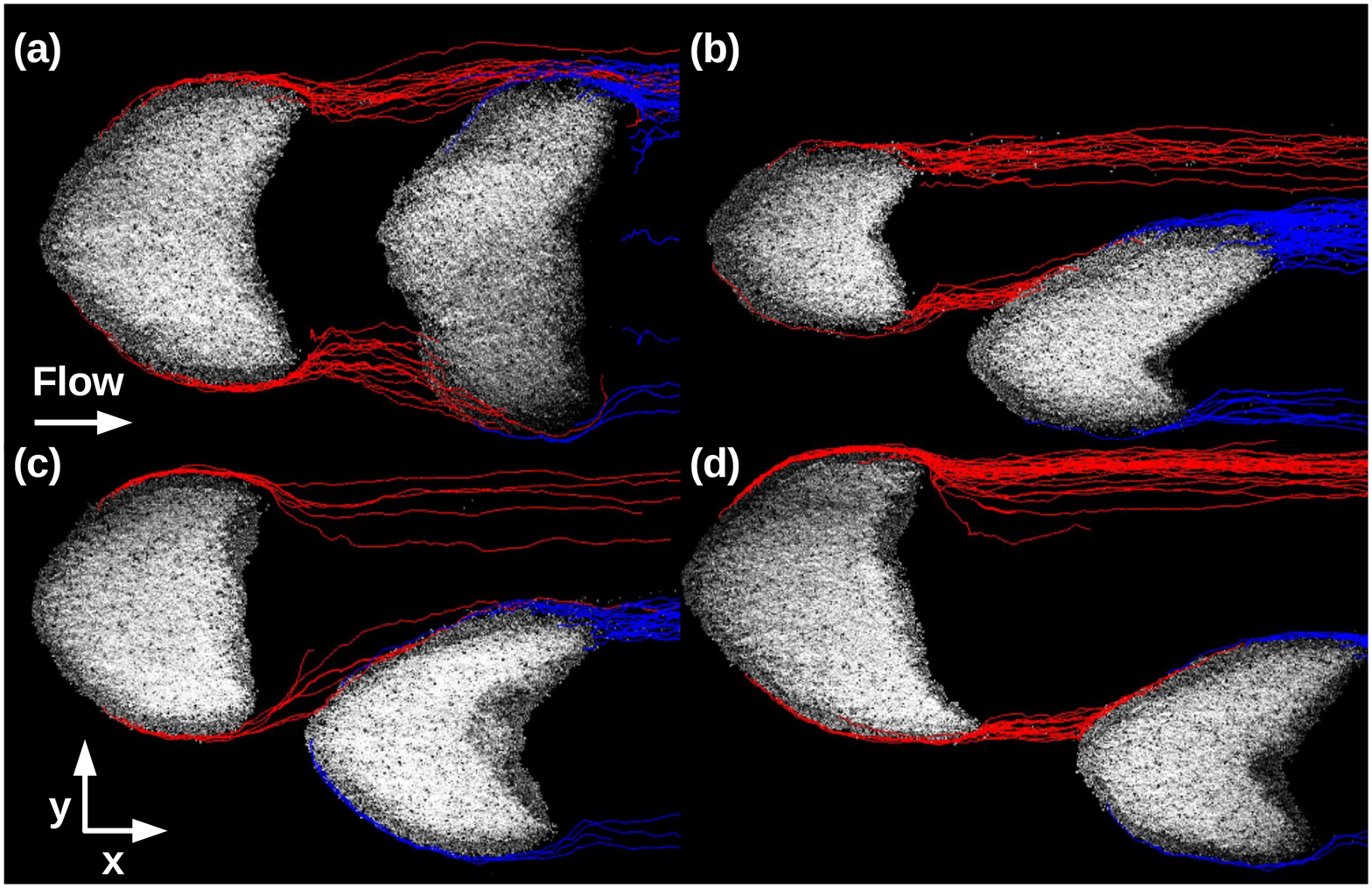}
\caption{Trajectories of some grains at two different intervals for the chasing pattern. Figures (a) and (b) correspond to two different stages of the interaction for the aligned case, and figures (c) and (d) to two different stages for the off-centered case. Red lines correspond to grains leaving the upstream dune and blue lines to grains leaving the downstream one.}
	\label{fig:trajectories_chasing}
\end{figure}

\begin{figure}
\centering
\includegraphics[width=0.8\linewidth]{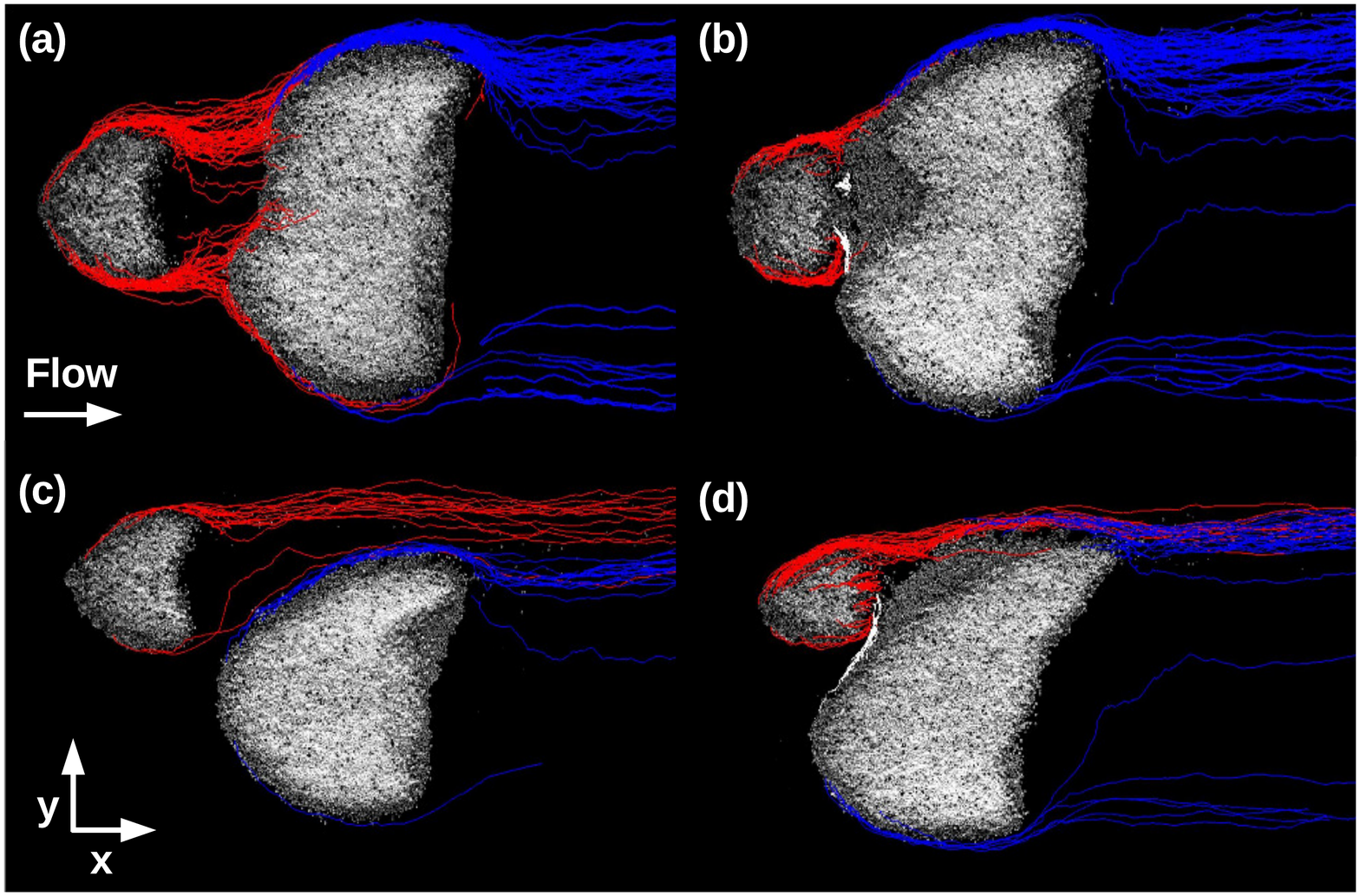}
\caption{Trajectories of some grains at two different intervals for the merging pattern. Figures (a) and (b) correspond to two different stages of the interaction for the aligned case, and figures (c) and (d) to two different stages for the off-centered case. Red lines correspond to grains leaving the upstream dune, blue lines to grains leaving the downstream one, and white lines to grains migrating from the downstream bedform to the upstream one.}
	\label{fig:trajectories_merging}
\end{figure}

\begin{figure}
\centering
\includegraphics[width=0.8\linewidth]{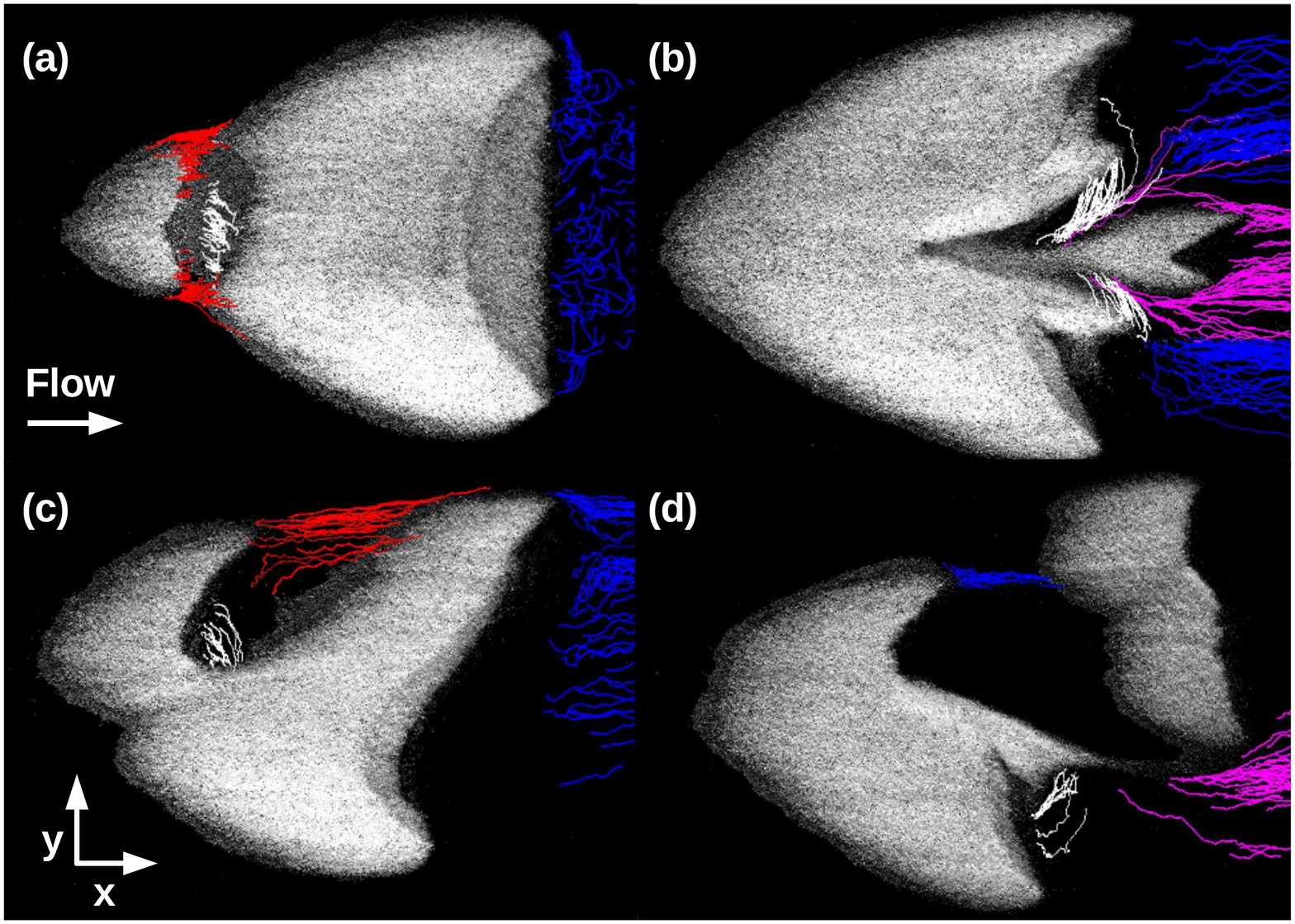}
\caption{Trajectories of some grains at two different intervals for the exchange pattern. Figures (a) and (b) correspond to two different stages of the interaction for the aligned case, and figures (c) and (d) to two different stages for the off-centered case. Red lines correspond to grains leaving the upstream dune, blue lines to grains leaving the downstream one, white lines to grains migrating from the downstream bedform to the upstream one, and magenta lines to grains leaving the new bedform.}
	\label{fig:trajectories_exchange}
\end{figure}

\begin{figure}
\centering
\includegraphics[width=0.8\linewidth]{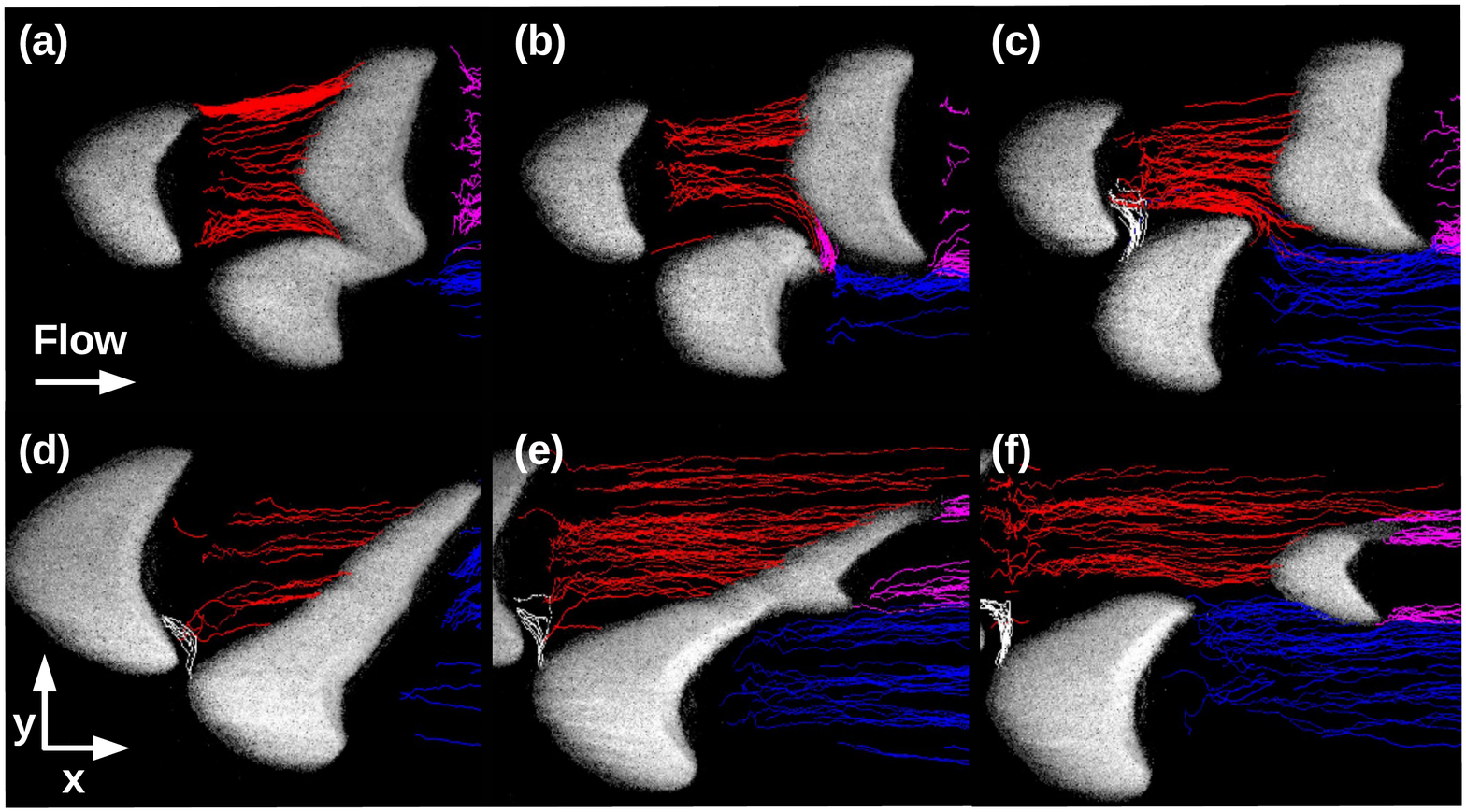}
\caption{Trajectories of some grains at three different intervals for the fragmentation-chasing pattern. Figures (a), (b) and (c) correspond to three different stages of the interaction for the aligned case, and figures (d), (e) and (f) to three different stages for the off-centered case. Red lines correspond to grains leaving the upstream dune, blue lines to grains leaving the downstream one, white lines to grains migrating from the downstream bedform to the upstream one, and magenta lines to grains leaving a new bedform.}
	\label{fig:trajectories_FragChasing}
\end{figure}

\begin{figure}
\centering
\includegraphics[width=0.8\linewidth]{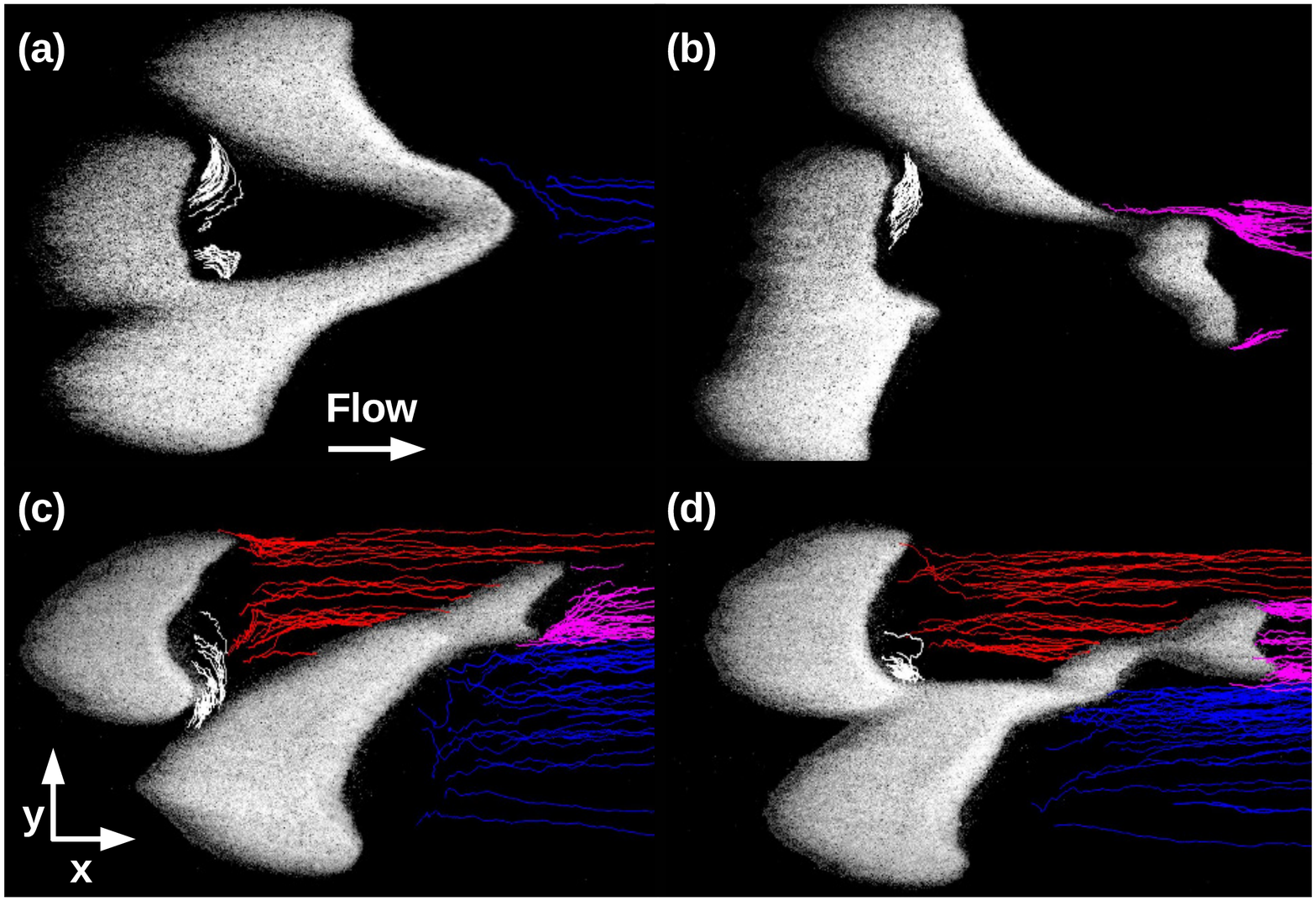}
\caption{Trajectories of some grains at two different intervals for the fragmentation-exchange pattern. Figures (a) and (b) correspond to two different stages of the interaction for the aligned case, and figures (c) and (d) to two different stages for the off-centered case. Red lines correspond to grains leaving the upstream dune, blue lines to grains leaving the downstream one, white lines to grains migrating from the downstream bedform to the upstream one, and magenta lines to grains leaving a new bedform.}
	\label{fig:trajectories_FragExchange}
\end{figure}

\begin{sloppypar}
We tracked moving grains during the interaction of barchans, and computed their trajectories. For these grains, the motion was mainly continuous, though with a stick-slip character, occurring directly over the channel wall (acrylic) and happening for a short time (that necessary for traveling from one barchan to the other). The trajectories of grains migrating from one dune to another, and also of grains leaving dunes and being entrained further downstream by the fluid, are of particular interest. Those trajectories reveal not only the masses exchanged between nearby dunes and lost by the entire system, but also details on how these exchanges and losses occur. Figures \ref{fig:trajectories_chasing} to \ref{fig:trajectories_FragExchange} show the trajectories of grains during different stages of barchan-barchan interactions, for all the five patterns in both aligned and off-centered configurations. Figures \ref{fig:trajectories_chasing} to \ref{fig:trajectories_FragExchange} correspond to the chasing, merging, exchange, fragmentation-chasing and fragmentation-exchange patterns \cite{Assis}, respectively, where subfigures on the top are related to aligned and on the bottom to off-centered cases. Red lines correspond to grains leaving the upstream (impact) dune, blue lines to grains leaving the downstream (target) dune, white lines to grains migrating from a downstream bedform to the upstream one, and magenta lines to grains leaving a new bedform. Whenever the bedforms are not the original impact and target barchans (case of white trajectories), the upstream bedform is considered as the one whose centroid is in an upstream position with respect to the other bedform. For the sake of clarity, the trajectories of a small portion of grains are plotted in Figures \ref{fig:trajectories_chasing} to \ref{fig:trajectories_FragExchange} (in average, 45\% of trajectories that took place during approximately 9 s were plotted, but percentages vary from 5\% to 100\% depending on the case), all trajectory types being shown, however.
\end{sloppypar}
 
For the chasing pattern (Figure \ref{fig:trajectories_chasing}), the wake of the upstream barchan strongly affects the downstream one \cite{Bristow, Bristow2, Bristow3}, the downstream barchan being strongly eroded and, due to small asymmetries, becoming eventually off-centered even in the aligned case.  We observe a large number of grains leaving the downstream barchan (in the aligned case, once dunes become off-centered), and the asymmetry of horns increases due to grains received asymmetrically from the upstream barchan. With both cases being eventually in an off-centered configuration, only grains from one of the horns of the upstream barchan reach the downstream one, and part of them simply go around the downstream barchan. At that stage (Figures \ref{fig:trajectories_chasing}b and \ref{fig:trajectories_chasing}d), we measured that approximately 25\% of grains leaving the upstream barchan go over the downstream dune (24\% in the aligned and 28\% in the off-centered case, which correspond to mass flow rates of 4.60 $\times$ 10$^{-3}$ and 1.12 $\times$ 10$^{-3}$ g/s, respectively), and that 7\% go around it and 69\% are directly entrained further downstream in the aligned case (mass flow rates of 1.34 $\times$ 10$^{-3}$ and 1.32 $\times$ 10$^{-2}$ g/s, respectively), while 44\% go around the downstream barchan and 28\% are directly entrained further downstream in the off-centered case (mass flow rates of 1.75 $\times$ 10$^{-3}$ and 1.12 $\times$ 10$^{-3}$ g/s, respectively). In addition, we computed the difference between grains received and lost by the downstream barchan (still at the late stage) and found that it reaches deficits of 18\% and 33\% in the aligned and off-centered cases, respectively (net flow rates of 3.45 $\times$ 10$^{-3}$ and 1.32 $\times$ 10$^{-3}$ g/s, respectively). The measured deficits corroborate the size decrease of the downstream barchan in the chasing pattern.

For the merging pattern (Figure \ref{fig:trajectories_merging}), we observe some differences between the aligned and off-centered cases. At the initial stage of the aligned case, a great part of grains leaving the upstream dune reaches the downstream one (92\% of grains are incorporated by the downstream bedform, which corresponds to 3.69 $\times$ 10$^{-3}$ g/s), deforming the downstream barchan into a barchanoid form. At a later stage, when dunes are almost colliding, the recirculation region in the wake of the upstream barchan carries grains from the downstream bedform to the upstream one, eroding the toe of the downstream bedform and forming a monolayer carpet between dunes before merging occurs. In the off-centered case, the main differences are that a much smaller number of grains leaving the upstream dune at the initial stage reaches the downstream one (only 1\% of them, corresponding to 3.65 $\times$ 10$^{-5}$ g/s), and that, before merging occurs, the recirculation region of the upstream barchan does not strongly erode the leading edge of the downstream dune, forming only the monolayer carpet.

During the initial stages, the behaviors of the exchange pattern in aligned and off-centered configurations (Figures \ref{fig:trajectories_exchange}a and \ref{fig:trajectories_exchange}c) are similar to those of the merging pattern (Figures \ref{fig:trajectories_merging}a and \ref{fig:trajectories_merging}c), the main difference being that grains leave the target barchan along all the lee face, instead of only through the horns. In the off-centered case, grains do not leave the target barchan from its horn farther from the upstream dune. This transverse distribution of the granular flux is caused by disturbances in the fluid flow (due to the upstream barchan), the flux of parting grains being concentrated out of horns in the case of an isolated barchan, as shown in the supporting information. We note that in Figure \ref{fig:trajectories_exchange}a the field of view does not allow us to follow the parting grains further downstream. However, during the tests we noticed that, indeed, a part of grains is entrained further downstream from the avalanche/lee face. After collision has taken place, the perturbation caused by the impacting barchan leads the resulting bedform to eject a new barchan. Along this text, we refer sometimes to the resulting (merged) and ejected bedforms as \textit{parent} and \textit{baby} barchans, respectively. In the aligned case, the new barchan is ejected from a central position at the lee face, and we observe that a considerable part of grains migrate from the new ejected barchan toward the upstream bedform (22\% of the grains that leave the new barchan, corresponding to 1.02 $\times$ 10$^{-3}$ g/s), forming two branches connecting, during a certain period, both dunes. In the off-centered case, the new barchan is ejected from one of the horns, and grains do not migrate from the ejected barchan toward the upstream dune. Instead, the ejected barchan continues receiving grains from the upstream dune.

In the fragmentation-chasing pattern (Figure \ref{fig:trajectories_FragChasing}), the perturbation caused by the wake of the upstream barchan is so strong that it splits the downstream dune into two smaller barchans. In both the aligned and off-centered cases, the downstream dune receives grains from the upstream barchan, but loses a larger quantity of grains (reaching deficits of 73\% and 19\% in the aligned and off-centered cases, respectively, which correspond to 3.92 $\times$ 10$^{-4}$ and 7.92 $\times$ 10$^{-5}$ g/s). Because of the perturbation of the fluid flow, grains leave barchans along the lee face, the exception being the smaller of split barchans of the off-centered case (perhaps also in the aligned case, but we did not follow it farther in the channel given the limitations of our traveling system). Once divided, the new barchans travel faster than the upstream one and they do not collide. However, one of the new barchans remains for some time close to the upstream one and some of its grains migrate toward the latter, entrained by the recirculation region.

The fragmentation-exchange pattern (Figure \ref{fig:trajectories_FragExchange}) is roughly similar to the fragmentation-chasing one, the main difference being that the impact barchan collides with one of the split bedforms. During the collision process, some grains are entrained from the downstream dune toward the impact barchan by the wake of the latter. In the aligned case, grains entrained further downstream leave barchans through one of the horns (mainly through the shared horn), while in the off-centered case parting grains are distributed along the lee face. Around 45\% of the grains leaving the downstream bedforms migrate to the impact barchan in the aligned case (46\% and 42\% in Figures \ref{fig:trajectories_FragExchange}a and \ref{fig:trajectories_FragExchange}b, respectively, which correspond to 6.55 $\times$ 10$^{-5}$ and 4.31 $\times$ 10$^{-5}$ g/s), while the percentages are 70\% and 5\% for the two stages of the off-centered case shown in Figures \ref{fig:trajectories_FragExchange}c and \ref{fig:trajectories_FragExchange}d, respectively (corresponding to 1.69 $\times$ 10$^{-3}$ and 8.05 $\times$ 10$^{-5}$ g/s). In the aligned case these percentages consider both split bedforms, while those for the off-centered case consider only grains from the split bedform closer to the impact barchan. The high percentage found in the approaching of barchans in the off-centered case (Figure \ref{fig:trajectories_FragExchange}c) reflects the formation of a granular bridge between them, which, once formed, unite both barchans with the consequent decrease in grains entrained toward the upstream dune (Figure \ref{fig:trajectories_FragExchange}d).

The chasing and fragmentation-chasing patterns are, perhaps, the three-dimensional equivalent of the dune-dune repulsion identified by \citeA{Bacik} in a narrow Couette-type circular channel, where the wake of the upstream bedform intensifies erosion on the downstream one, increasing the celerity of the latter. However, different from \citeA{Bacik}, our channel is relatively large, producing dune-dune repulsion cases where barchans become off-centered and split (in addition to the collision cases).

A table summarizing the percentages of grains exchanged between dunes, the total number of moving grains and the considered time interval is available in the supporting information. With that, we can estimate the overall transport of grains in the inter-dune space (migrating from one dune to another or being entrained further downstream), which is also presented in the supporting information (in terms of mass flow rates). In addition, trajectories of grains leaving an isolated subaqueous barchan are also available in the supporting information, from which we can observe that all grains leave the dune through their horns (a great part of them coming from upstream regions and going around the dune before reaching the horns, as shown by \citeA{Alvarez3} and \citeA{Alvarez4}).

\subsection{\label{sec:Res_lengths} Lengths and velocities of exchanged grains}

\begin{figure}
\centering
\includegraphics[width=0.85\linewidth]{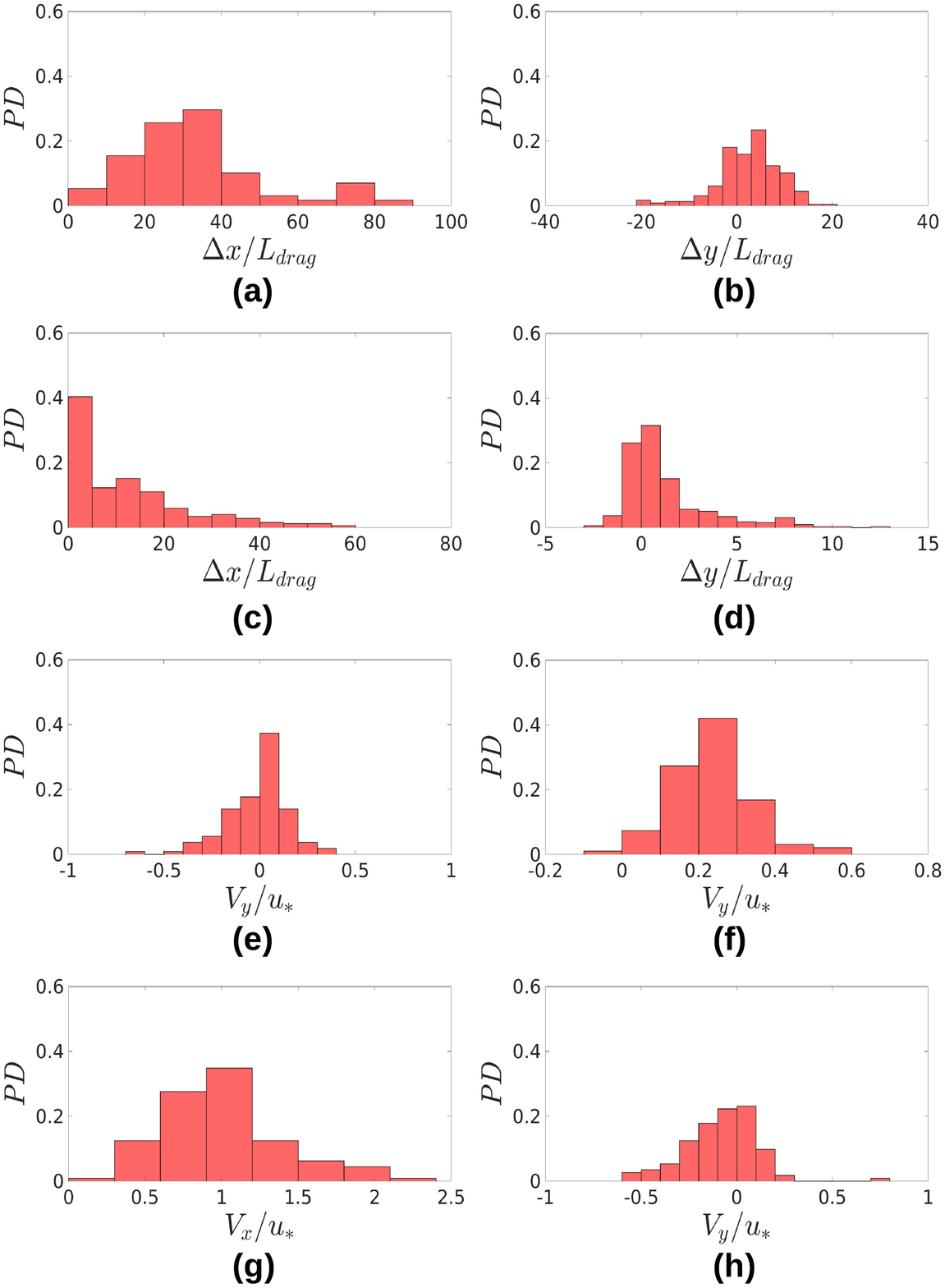}
\caption{PDs of total distances traveled by grains in longitudinal and transverse directions, $\Delta x$ and $\Delta y$, respectively, normalized by $L_{drag}$, and PDs of time-averaged velocities in the longitudinal and transverse directions, $V_x$ and $V_y$, respectively, normalized by $u_*$ (values of $u_*$ are available in the supporting information). Figures (a) and (b) correspond to red trajectories in Figure \ref{fig:trajectories_merging}a, with mean values of $\Delta x / L_{drag}$ = 33.4 and $\Delta y / L_{drag}$ = 2.4, and standard deviations of, respectively, 18.2 and 6.7$L_{drag}$. Figures (c) and (d) correspond to red trajectories in Figure \ref{fig:trajectories_merging}d, with mean values of $\Delta x / L_{drag}$ = 12.6 and $\Delta y / L_{drag}$ = 1.3, and standard deviations of, respectively, 12.7 and 2.4. Figures (e) and (f) correspond to red trajectories in Figures \ref{fig:trajectories_chasing}a and \ref{fig:trajectories_chasing}b with mean values of $V_y / u_*$ equal to -0.01 and 0.23, and standard deviations of 0.16 and 0.10, respectively. Figures (g) and (h) correspond to red trajectories in Figure \ref{fig:trajectories_FragChasing}c (grains leaving the impact barchan along its lee face), with mean values of $V_x / u_*$ = 1.01 and $V_y / u_*$ = -0.08, and standard deviations of 0.39 and 0.19, respectively.}
	\label{fig:PDFs_exchanged_grains}
\end{figure}

\begin{figure}
\centering
\includegraphics[width=0.85\linewidth]{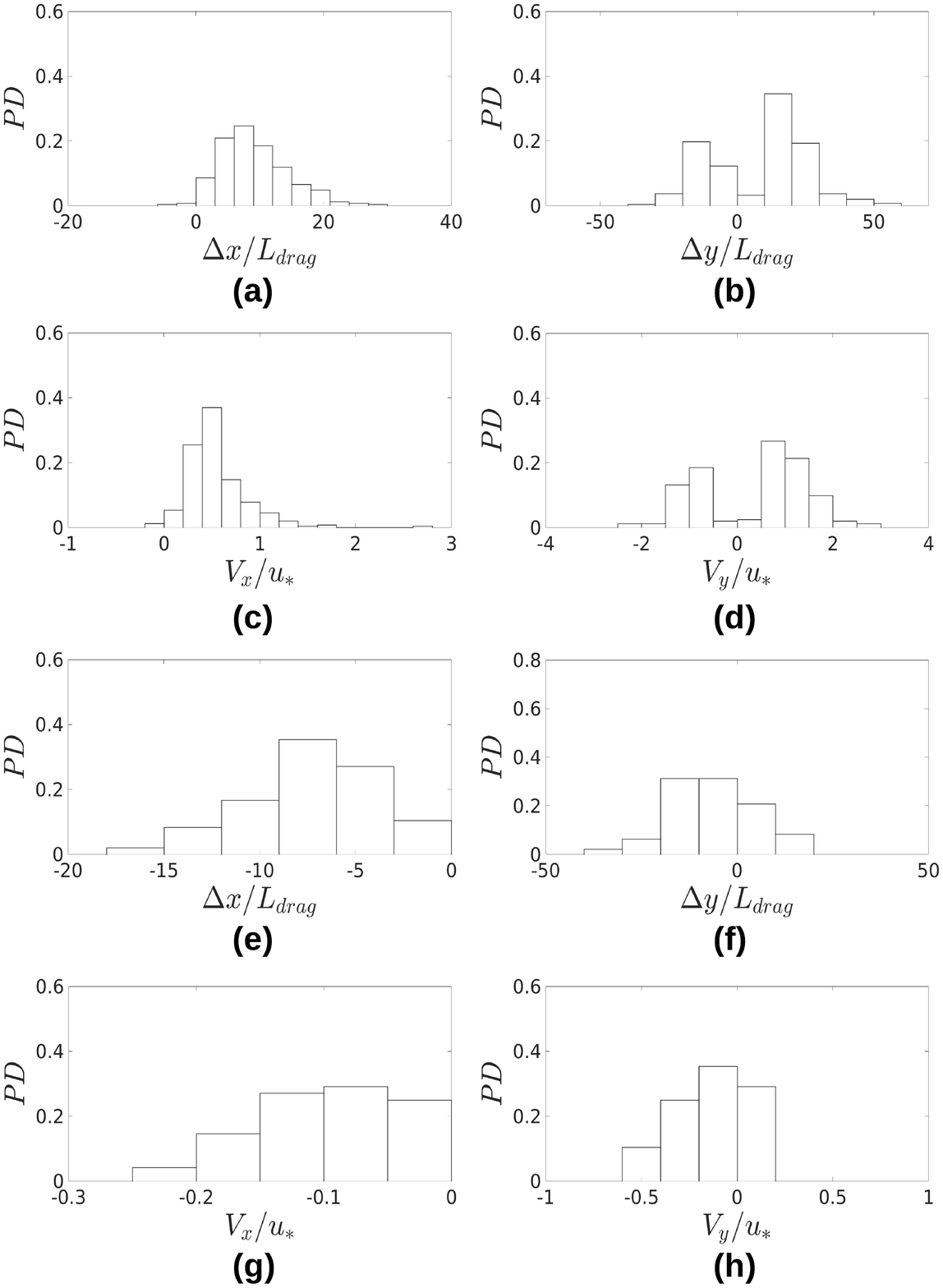}
\caption{PDs of total distances traveled by grains in longitudinal and transverse directions, $\Delta x$ and $\Delta y$, respectively, normalized by $L_{drag}$, and PDs of time-averaged velocities in the longitudinal and transverse directions, $V_x$ and $V_y$, respectively, normalized by $u_*$ (values of $u_*$ are available in the supporting information). Figures (a) to (d) correspond to white trajectories in Figure \ref{fig:trajectories_exchange}b, with mean values for lenghts of $\Delta x / L_{drag}$ = 9.1 and $\Delta y / L_{drag}$ = 8.0 and standard deviations of, respectively, 5.4 and 17.8, and mean values for velocities of $V_x / u_*$ = 0.55 and $V_y / u_*$ = 0.37, and standard deviations of 0.31 and 1.13, respectively. Figures (e) to (h) correspond to white trajectories in Figure \ref{fig:trajectories_FragExchange}a, with mean values for lengths of $\Delta x / L_{drag}$ = -7.2 and $\Delta y / L_{drag}$ = -6.3 and standard deviations of, rescpectively, 3.5 and 11.4, and mean values for velocities of $V_x / u_*$ = -0.10 and $V_y / u_*$ = -0.14, and standard deviations of 0.06 and 0.18, respectively.}
	\label{fig:PDFs_exchanged_grains1}
\end{figure}

Based on grain trajectories, we identified, for the characteristic routes distinguished in Subsection \ref{sec:Res_trajectories}, typical lengths and velocities of grains migrating from one dune to another. For that, we computed the displacement lengths in the longitudinal and transverse directions, $\Delta x$ and $\Delta y$, respectively, as well as the time-averaged velocities in the longitudinal and transverse directions, $V_x$ and $V_y$, respectively, of exchanged particles. Displacements were computed as the differences between the final and initial positions of each grain from their departure from one dune until reaching another one, and average velocities as the mean values for each grain during its migration. We then plotted their respective probability distributions (PDs) by considering all tracked particles, and present some of them in Figures \ref{fig:PDFs_exchanged_grains} and \ref{fig:PDFs_exchanged_grains1} (other PDs are available in the supporting information, including those for an isolated barchan). While the velocities were normalized by $u_*$, which is a characteristic velocity at the grain scale, lengths were normalized by $L_{drag}$ = $\rho_s \rho^{-1}d$ \cite{Hersen_1}. Although the saturation length $L_s$ proposed by \citeA{Pahtz_1} is the proper scale for the response of a granular bed to flow changes, and, therefore, for erosion and deposition and minimum bedform scales, we use $L_{drag}$ for normalizations. $L_s$ takes into account the forces controlling grain and fluid relaxation, both for gases and liquids, incorporating mechanisms not present in $L_{drag}$. However, $L_{drag}$ is a length scale of inertial nature with a simple expression proposed by \citeA{Hersen_1}, being a reasonable scaling for dune lengths over 5 orders of magnitude \cite{Claudin_Andreotti}.

Figure \ref{fig:PDFs_exchanged_grains} presents PDs of grains migrating from upstream to downstream barchans, and Figure \ref{fig:PDFs_exchanged_grains1} of those migrating from downstream to upstream bedforms. Distances and velocities for the remaining cases are similar to those shown in Figures \ref{fig:PDFs_exchanged_grains} and \ref{fig:PDFs_exchanged_grains1}, and some of them are presented in the supporting information, as well as those for a single barchan. In general, mean values of traveled distances in the longitudinal direction are proportional to the longitudinal separation between barchans, while in the transverse direction they are proportional to the transverse offset between bedforms. Because exchanged grains move directly over the channel wall (acrylic) when traveling from one barchan to another, defining an area over which they move is more difficult than for grains moving over a thick granular bed. However, distributions of $\Delta x$ and $\Delta y$ can be used to estimate the area swept by the tracked grains: for grains moving in the longitudinal direction, that area is proportional to $\Delta x$ multiplied by 2 times the standard deviation of $\Delta y$, and the contrary (in terms of $x$ and $y$) for grains moving in the transverse direction. Since we performed Lagrangian tracking, another area of interest is the cross-sectional area crossed by the followed particles. Because the exchanged grains roll directly over the channel wall, the height of that area is proportional to the grain diameter, while its width is proportional to the standard deviations of $\Delta y$ or $\Delta x$ for grains moving in the longitudinal or transverse directions, respectively. Therefore, bedload fluxes can be estimated as the mass flow rates divided by the corresponding cross-sectional areas. Concerning specifically the values of $\Delta y / L_{drag}$ measured for barchan-barchan interactions, they have mean values and standard deviations higher than those for the single dune (that has mean average and standard deviation of -0.62 and 3.16, respectively). In particular, for cases where channeling is present (red trajectories in Figures \ref{fig:trajectories_chasing}b, \ref{fig:trajectories_merging}c, \ref{fig:trajectories_merging}d and \ref{fig:trajectories_FragChasing}c, for example), $\Delta y / L_{drag}$ reaches values one or two orders of magnitude higher than those for the single dune, indicating a strong deflection in the trajectories of grains (values available in the supporting information).

For the longitudinal component of velocities, mean values are mostly positive but can be negative when grains are entrained by the recirculation region of the upstream dune, which happens in some cases when two bedforms are very close, almost touching each other. In the main, they are one order of magnitude smaller than the undisturbed shear velocity over the channel wall (reference value), $u_*$, but in some regions where the fluid flow is locally accelerated and/or has its turbulence level increased $V_x$ reaches values of the same order of magnitude of $u_*$ (same order of magnitude found for grains leaving an isolated barchan, shown in the supporting information). For the transverse component, mean values tend to zero for aligned bedforms, due to symmetry, and deviate from zero for off-centered bedforms (for reference, values for grains leaving an isolated barchan have an average of the order of 10$^{-3}u_*$). For some aligned bedfoms, such as during the ejection of a baby barchan in the aligned-exchange configuration, distributions of transverse displacements and velocities are bimodal and roughly symmetrical around a zero mean. In particular, we found that the grains migrating from the baby barchan toward the parent dune in the aligned-exchange configuration move downstream, with mean longitudinal distances of approximately 10$L_{drag}$ and transverse displacements of approximately 15$L_{drag}$. For the chasing pattern in aligned configuration, mean values of $V_y/u_*$ deviate from approximately zero toward other values (from -0.01 to 0.23 in the case of Figures \ref{fig:PDFs_exchanged_grains}e and \ref{fig:PDFs_exchanged_grains}f), this being a consequence of wake interactions, including channeling, that lead the aligned configuration toward an off-centered one. In all interacting cases, wake effects and small asymmetries are present, the mean value of $V_y/u_*$ for the single dune being at least one order of magnitude smaller when compared to all cases ($V_y/u_*$ = -0.006 for the single dune).

\subsection{\label{sec:Granular_spreading} Spreading after collision}

Having analyzed in Subsections \ref{sec:Res_trajectories} and \ref{sec:Res_lengths} the motion of grains between bedforms, we investigate now the motion, after collision has taken place, of grains originally in the impact barchan. For that, we present data at both the barchan and grain scales. At the barchan scale, some of the images obtained by \citeA{Assis} are now further treated for measuring the spreading of the impacting bedform based on the evolution of its area over the target barchan. At the grain scale, we determine, from new movies, typical trajectories of individual grains by tracking their motion once collision has occurred.

\begin{figure}
\centering
\includegraphics[width=0.85\linewidth]{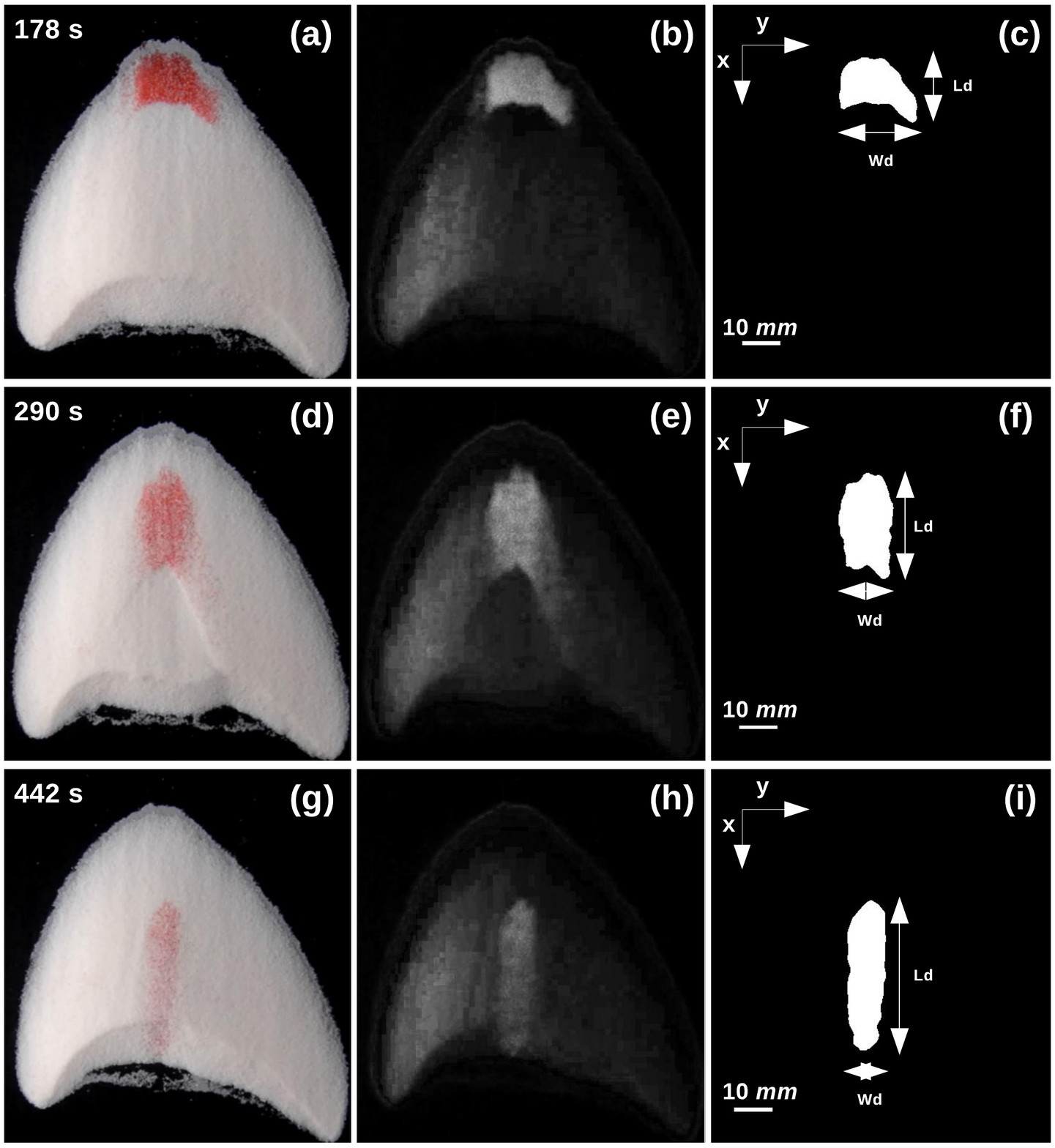}
\caption{Distribution of grains from the impact barchan over the target one during a merging process in the aligned case. Red (clear in figures b, e and h) grains come from the impact barchan and white (darker in figures b, e and h) grains are from the target one. From top to bottom, figures correspond to different instants (shown in figures), and from left to right figures correspond to raw, grayscale and binary images. $L_d$ is the length and $W_d$ the width of the structure formed with grains from the impact barchan.}
	\label{fig:diffusion_merging}
\end{figure}

\begin{figure}
\centering
\includegraphics[width=0.85\linewidth]{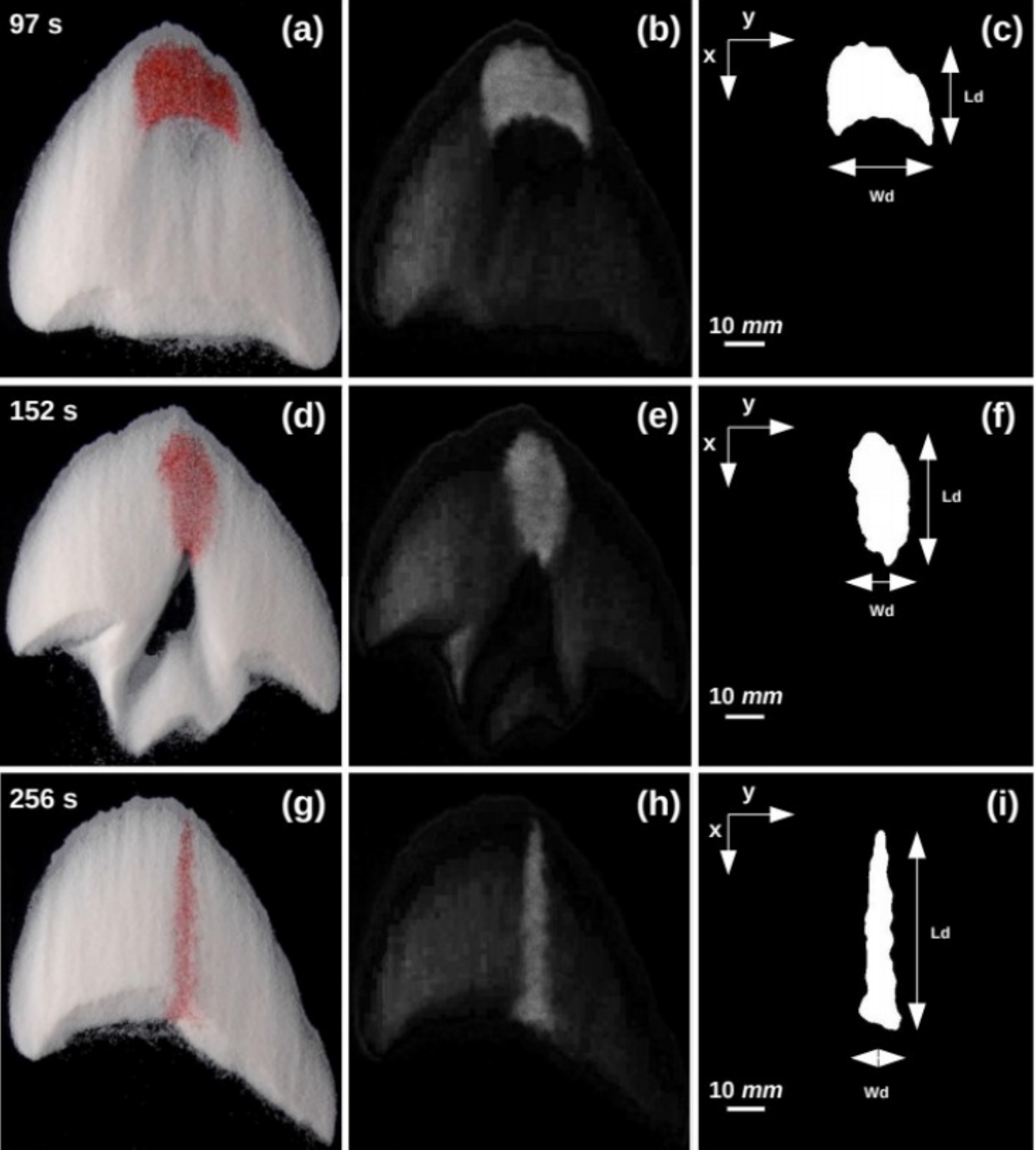}
\caption{Distribution of grains from the impact barchan over the target one during an exchange process in the aligned case. Red (clear in figures b, e and h) grains come from the impact barchan and white (darker in figures b, e and h) grains are from the target one. From top to bottom, figures correspond to different instants (shown in figures), and from left to right figures correspond to raw, grayscale and binary images. $L_d$ is the length and $W_d$ the width of the structure formed with grains from the impact barchan.}
	\label{fig:diffusion_exchange}
\end{figure}

\begin{figure}
\centering
\includegraphics[width=0.85\linewidth]{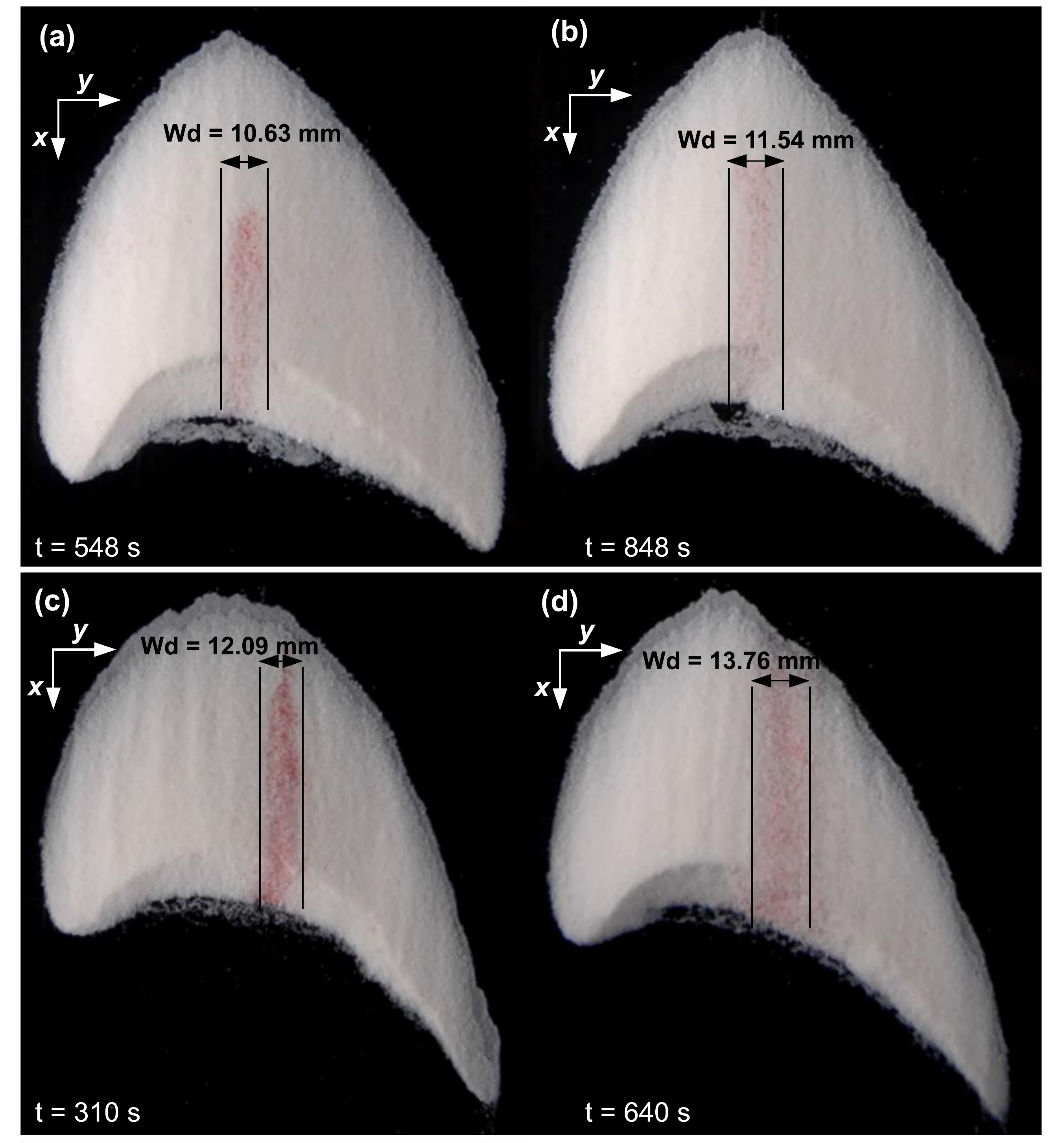}
\caption{Second stage of the spreading of grains from the impact dune over the target one during (a) and (b) merging and (c) and (d) exchange processes. Red grains come from the impact barchan. Times and lengths are shown in the figure.}
	\label{fig:diffusion_bedform_scale}
\end{figure}

We notice two distinct stages in the evolution of the impacting bedform in the merging and exchange cases. The first stage corresponds to a barchan shape being stretched and becoming a longitudinal stripe, while in the second one the stripe widens slowly along time. Both stages can be observed in Figures \ref{fig:diffusion_merging} to \ref{fig:diffusion_bedform_scale}, which show grains from the impact barchan over the target one, and also in the supporting information, which shows the width of the longitudinal stripe $W_d$ as a function of time. We note that a two-stage adaptation of dunes to a change in the flow conditions has already been proposed by \citeA{Fischer}. Based on 2D simulations using a minimum model \cite{Kroy_A, Kroy_C}, they showed that dunes are unstable solutions: once disturbed, a first stage that corresponds to an adaptation of shape to the new unstable conditions takes place, followed by a second stage where mass changes along these new conditions. This seems to bear similarities with the spreading of the impact barchan after collision takes place. However, while the initial flattening can in part be explained as an adaptation to new conditions, sharing perhaps similarities with \citeA{Fischer}, the transverse diffusion does not seem related to numerical results based on 2D bedforms since the widening of the longitudinal stripe results from a diffusion-like mechanism over the resulting barchan, not acting directly on the same bedform (the impact barchan, which has flattened in the first stage).

For the first stage, Figures \ref{fig:diffusion_merging} and \ref{fig:diffusion_exchange} show how grains originally in the impact barchan spread over the target one during merging and exchange processes, respectively, in the aligned case. Red (clear) regions correspond to grains from the impact barchan and white (darker) to grains from the target one, and different instants are shown from top to bottom. We observe that, while the impacting bedform is deformed into a longitudinal stripe, its wake disturbs the surface of the target barchan. In the case of the exchange pattern, the perturbation is strong enough to eject a new barchan that does not contain grains from the impact dune, while in the merging pattern the perturbation is attenuated. The difference in patterns may be related to the lengths of surface waves, that propagate faster than the resulting barchan and, if not attenuated, can eject a new barchan by calving \cite{Elbelrhiti, Worman}. \citeA{Claudin_Andreotti} showed that the minimum wavelength for subaqueous waves is approximately 20 mm, meaning that only waves longer than that value persist and produce calving. If we consider the initial values of $W_d$ as the typical length of the impact barchan, then $W_d$ $<$ 20 mm for the merging and $W_d$ $>$ 20 mm for the exchanges patterns, in accordance with \citeA{Claudin_Andreotti} (see the supporting information for the evolutions of $W_d$ along time). The ejection of a baby barchan seems thus in accordance with the calving mechanism, as proposed by \citeA{Elbelrhiti} and \citeA{Worman}.

Figure \ref{fig:diffusion_bedform_scale} presents the second stage of the deformation of the impacting bedform, Figures \ref{fig:diffusion_bedform_scale}a and \ref{fig:diffusion_bedform_scale}b corresponding to a merging process and Figures \ref{fig:diffusion_bedform_scale}c and \ref{fig:diffusion_bedform_scale}d to an exchange process. We observe that the longitudinal stripe widens slowly along time, in what resembles a diffusion process, taking 300 s to widen 0.91 mm in the merging case and 330 s to widen 1.67 mm in the exchange case. The corresponding expansion (widening) velocities are, respectively, 3 $\times$ 10$^{-6}$ and 5 $\times$ 10$^{-6}$ m/s, which correspond to 2 $\times$ 10$^{-4}$ and 3 $\times$ 10$^{-4}$$u_*$, while grain velocities are much larger, of the order of 10$^{-1}$$u_*$ (as shown next). Because the main flow is in the longitudinal direction, we conjecture that the widening of the longitudinal stripe is caused by the erratic trajectories of grains, which are, in addition, amplified in the transverse direction due to the lateral slopes of the bedform. Although not a pure diffusion in the strict sense, we describe next this widening processes as a diffusion-like mechanism given the resemblance. In order to investigate that, we followed individual grains during the stripe widening and computed their trajectories, displacement lengths and velocities.

\begin{figure}
\centering
\includegraphics[width=0.9\linewidth]{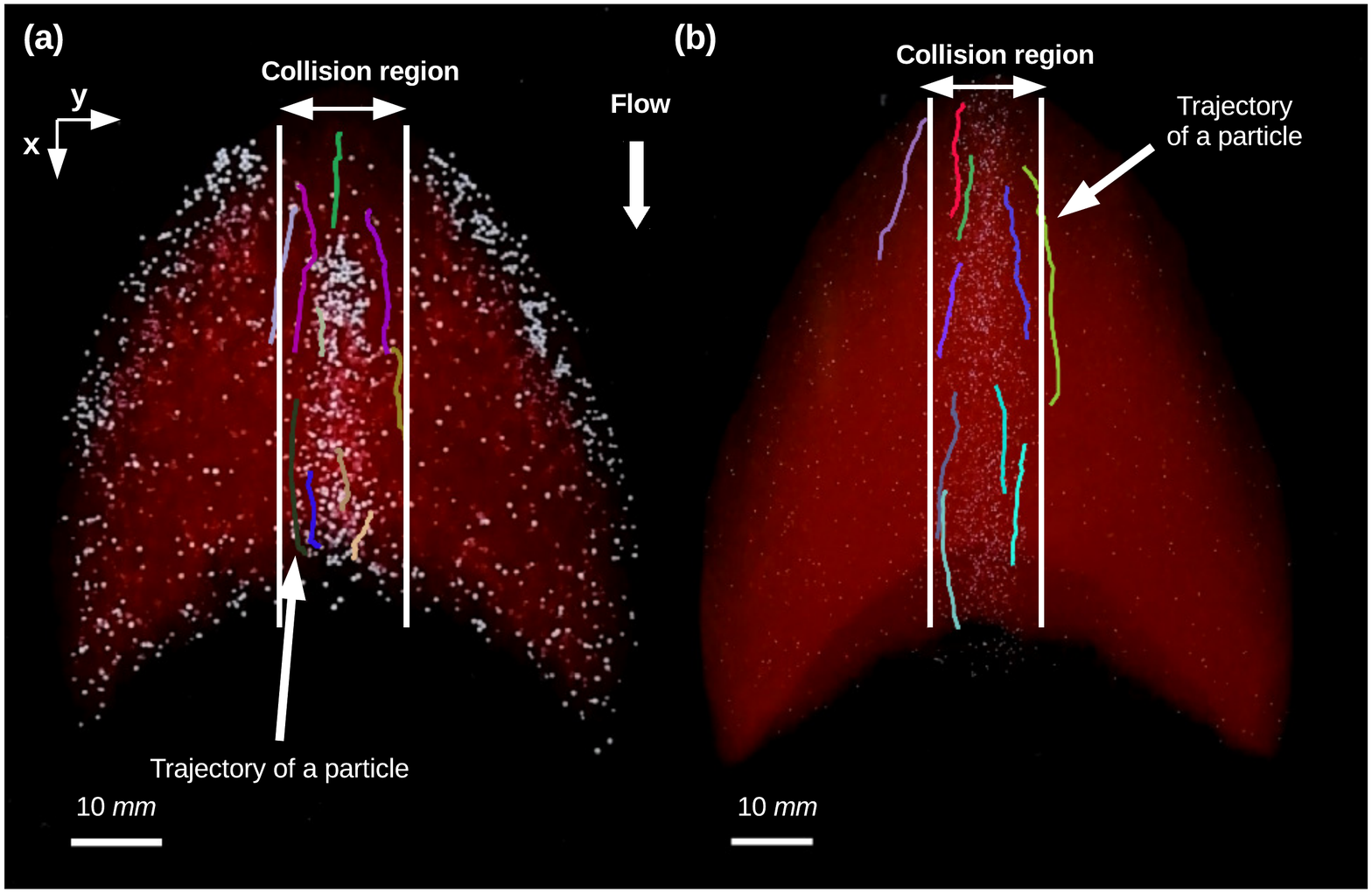}
\caption{Typical trajectories of individual grains from the impact barchan over the target one for (a) merging and (b) exchange.}
	\label{fig:diffusion_grain_scale}
\end{figure}

Figures \ref{fig:diffusion_grain_scale}a and \ref{fig:diffusion_grain_scale}b show some trajectories of grains (from the impact dune) over the target barchan for the merging and exchange processes, respectively. The motion of these grains was intermittent and occurred from a starting point until reaching the crest region. We observe a small transverse component that varies from grain to grain that contributes to the stripe widening. In order to scrutinize their relation, we computed mean values and standard deviations of displacements and velocities for a large amount of particles, obtaining diffusion-like measurements at the grain scale. The considered grains were those from the impact barchan that started moving over the target barchan at positions within a width equivalent to that of the impact dune (boundaries shown in Figure \ref{fig:diffusion_grain_scale}).

\begin{figure}
\centering
\includegraphics[width=0.85\linewidth]{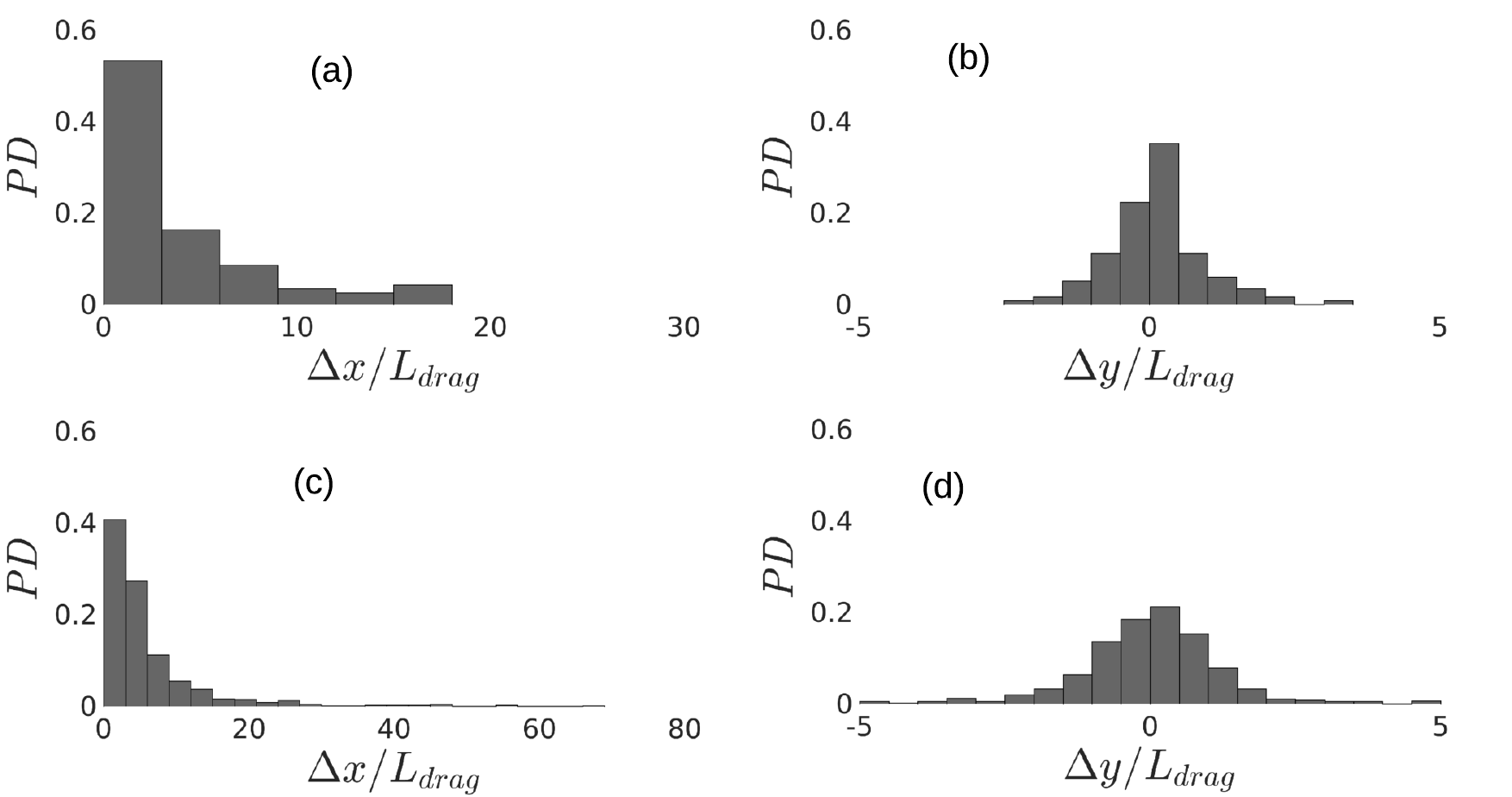}
\caption{PDs of total distances traveled by grains in: (a) and (c) the longitudinal direction and (b) and (d) the transverse direction, $\Delta x$ and $\Delta y$, respectively, normalized by $L_{drag}$. Figures (a) and (b) correspond to the merging and (c) and (d) to the exchange pattern.}
	\label{fig:diffusion_lengths_grains}
\end{figure}

Figure \ref{fig:diffusion_lengths_grains} presents PDs of total distances traveled by the followed grains in the longitudinal and transverse directions, $\Delta x$ (Figures \ref{fig:diffusion_lengths_grains}a and \ref{fig:diffusion_lengths_grains}c) and $\Delta y$ (Figures \ref{fig:diffusion_lengths_grains}b and \ref{fig:diffusion_lengths_grains}d), respectively. These distances correspond to the differences between the final and initial positions of each grain, and they are normalized by $L_{drag}$. Figures \ref{fig:diffusion_lengths_grains}a and \ref{fig:diffusion_lengths_grains}b correspond to the merging and Figures \ref{fig:diffusion_lengths_grains}c and \ref{fig:diffusion_lengths_grains}d to the exchange pattern. PDs of $\Delta x / L_{drag}$ have a decreasing distribution that seems exponential, but we prefer to not assert its form for the moment, however, given the relative small size of our samples. Longitudinal distances have average values of approximately 4 and 6$L_{drag}$ and RMS (root mean square) averages of 5 and 10$L_{drag}$ for the merging and exchange patterns, respectively. Distributions of $\Delta y / L_{drag}$ show a Gaussian-like behavior, peaked close to zero. Here again, we prefer to not assert the form of the distribution. Transverse distances traveled by the followed grains show average values of approximately 0.1 and -0.2$L_{drag}$, standard deviations of 0.8 and 1.5$L_{drag}$, and RMS averages of 0.8 and 1.5$L_{drag}$ for the merging and exchange patterns, respectively. These values show ensemble averages around zero with large dispersions, indicating that grains travel longitudinally with considerable deviations in the transverse direction that are symmetrical with respect to the longitudinal direction. This kind of trajectory spreads the longitudinal stripe in a way that resembles a diffusion mechanism.

\begin{figure}
\centering
\includegraphics[width=0.95\linewidth]{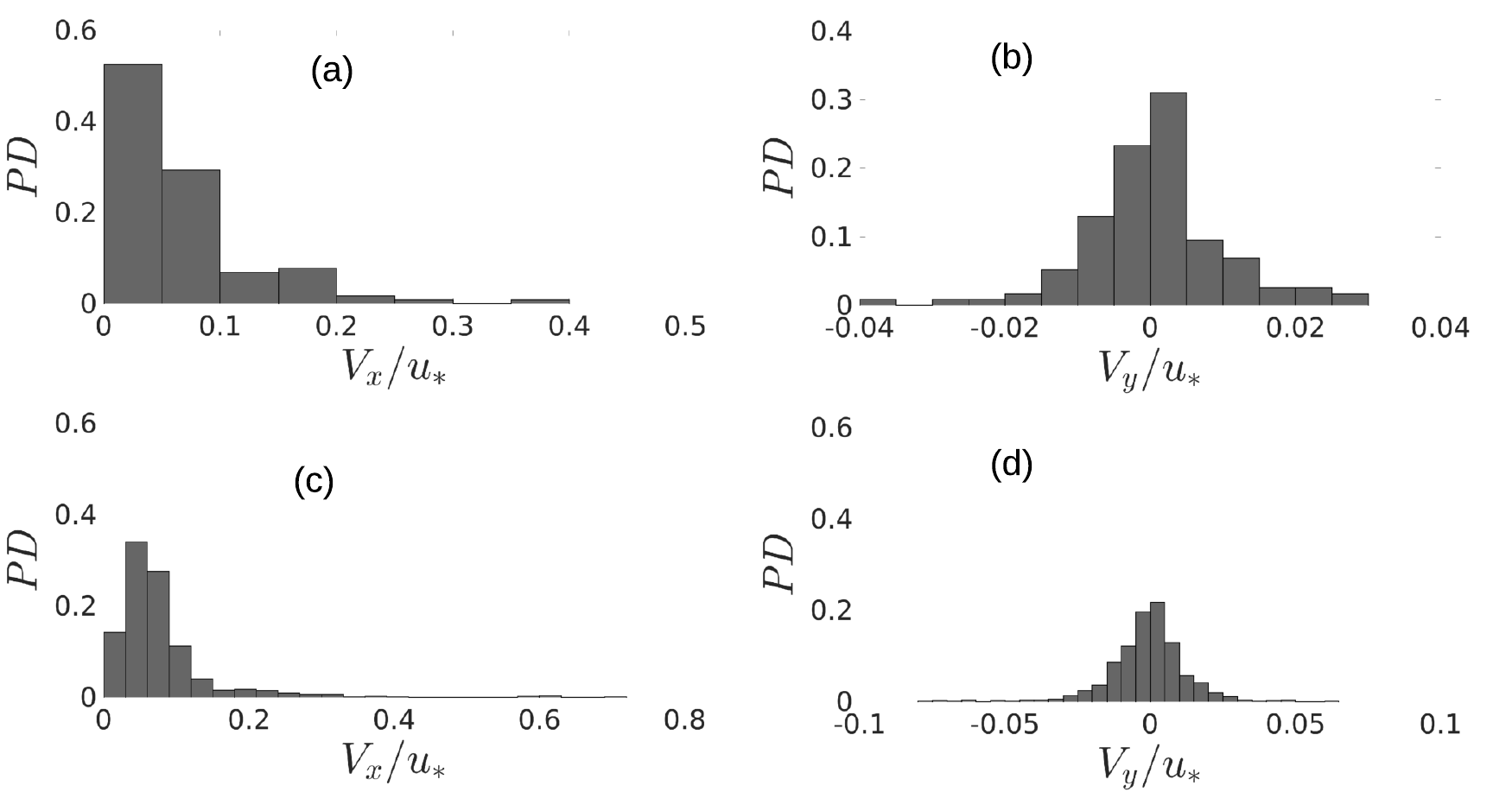}
\caption{PDs of time-averaged velocities in: (a) and (c) the longitudinal direction and (b) and (d) the transverse direction, $V_x$ and $V_y$, respectively, normalized by $u_*$. Figures (a) and (b) correspond to the merging and (c) and (d) to the exchange pattern.}
	\label{fig:diffusion_velocities_grains}
\end{figure}

Figure \ref{fig:diffusion_velocities_grains} presents mean velocities of grains during their trajectories in the longitudinal and transverse directions, $V_x$ (Figures \ref{fig:diffusion_velocities_grains}a and \ref{fig:diffusion_velocities_grains}c) and $V_y$ (Figures \ref{fig:diffusion_velocities_grains}b and \ref{fig:diffusion_velocities_grains}d), respectively, normalized by $u_*$. Each mean value in the PDs was computed as the time-averaged velocity of each grain during its displacement over the dune. PDs of $V_x$ show a decreasing distribution that is monotonic for the merging and non-monotonic for the exchange pattern. The mean velocities in the longitudinal direction are of the order of 0.1$u_*$: $V_x$ presents average values of 0.07 and 0.08$u_*$, standard deviations of 0.06 and 0.08$u_*$, and RMS averages of 0.09 and 0.11$u_*$ for the merging and exchange patterns, respectively. PDs of $V_y$ are peaked close to zero, and present average values of approximately  4.0 $\times$ 10$^{-4}$ and -1.5 $\times$ 10$^{-3}$$u_*$, standard deviations of approximately 1.0 $\times$ 10$^{-2}$ and 1.4 $\times$ 10$^{-2}$$u_*$, and RMS averages of 9.5 $\times$ 10$^{-3}$ and 1.4 $\times$ 10$^{-2}$$u_*$ for the merging and exchange patterns, respectively. As for $\Delta y$, transverse velocities present an ensemble average around zero with large dispersion, indicating motions in the transverse direction that are symmetrical with respect to the longitudinal direction. For each grain, we computed the RMS average of the transverse velocity, $Vy_{rms}$, during its trajectory over the barchan, and present the corresponding PDs in supporting information. From the RMS PDs, we find average values of 3 $\times$ 10$^{-4}$ and 7 $\times$ 10$^{-4}$$u_*$ (5 $\times$ 10$^{-6}$ and 11 $\times$ 10$^{-6}$ m/s) for the merging and exchange cases, respectively. These values are of the same order of magnitude of those obtained for the expansion of the longitudinal stripe. PDs of $\Delta x$, $\Delta y$, $V_x$, $V_y$ and $Vy_{rms}$ in dimensional form are available in the supporting information.

Finally, we computed the diffusion length $l_d$ = $\sigma_y^2 / (2\Delta x)$, where $\sigma_y$ is the standard deviation of the transverse displacement, as proposed by \citeA{Seizilles} for bedload over a plane bed, though in the present case grains move over a curved bed: they follow an upward slope along the symmetry line, with a varying lateral inclination from the symmetry line toward the flanks. We found $l_d/L_{drag}$ $\approx$ 0.10 and 0.20 (corresponding to $l_d/d$ $\approx$ 0.3 and 0.5) for the merging and exchange patterns, respectively. These values are one order of magnitude higher than that obtained by \citeA{Seizilles}, who found  $l_d/L_{drag}$ $\approx$ 0.012 (or $l_d/d$ $\approx$ 0.03). We believe that the lateral slope amplify the transverse component of the motion in subaqueous bedload, which has an erratic origin \cite{Seizilles}, improving significantly the transverse diffusion and increasing $l_d$ by one order of magnitude. For the upward slope in itself, we believe that it has no significant effect on $l_d$ since the diffusion-like mechanism occurs in the transverse direction.

\section{\label{sec:Conclu} Conclusions}

We investigated the motion of grains while two barchans interacted with each other by performing experiments in a water channel, recording images with high-speed and conventional cameras, and tracking bedforms and individual grains along images. We found typical trajectories of grains during barchan-barchan interactions, from which we determined the origin and destination of moving grains, the proportions of grains exchanged between barchans and lost by the entire system, the respective mass flow rates, and the typical lengths and velocities of grains following different paths. Among our findings, we showed that the approximate deficits of granular fluxes in the aligned and off-centered configurations reach, respectively, 20 and 30\% for the chasing and 60 and 20\% for the fragmentation-chasing patterns. Therefore, in these patterns the downstream bedforms decrease in size, moving faster and avoiding collision with the upstream dune. Interestingly, we found that during the ejection of a new barchan in the exchange pattern in aligned configuration, 20\% of grains leaving the baby barchan move toward the parent bedform, forming two granular branches that connect both dunes during a given period of time. In this particular case, we found that these grains move downstream, whereas in the exchange pattern in off-centered configuration there are no grains migrating from the baby barchan toward the parent dune, the same occurring in the fragmentation exchange case (for both aligned and off-centered configurations). In addition, we followed the bedforms after collision took place in the merging and exchange patterns, revealing an initial stage, where the impact barchan is stretched until becoming a longitudinal stripe, and a second stage where the stripe widens slowly. For the second stage, we followed grains originally in the impact barchan and showed that they spread with an erratic trajectory over the target dune, having transverse velocities that scale with the front velocity of the stripe and resembling a diffusion process. For these grains, we found a diffusion length $l_d$ of the order of 0.1$L_{drag}$, one order of magnitude higher than that obtained by \citeA{Seizilles} for subaqueous bedload over plane beds, and we conjecture that the lateral slopes of barchans amplify the transverse component of the erratic motion of grains.

In general, the following insights into the modeling of barchan-barchan interactions are gained from this study: (i) in certain cases, there are different trajectories for grains in the presence of an upstream perturbation (caused by the upstream barchan), with grains not being entrained further downstream from certain horns and/or leaving the target barchan along the lee face (instead of through the horns), for instance; (ii) the knowledge of how grains are exchanged between barchans, at different phases of the interaction patterns; (iii) the identification of a diffusion-like mechanism after collision has taken place, and a corresponding diffusion length. These results represent a step toward understanding the barchan coarsening and division, size selection, and variability of barchanoid shapes found in water, air, and other planetary environments.

\acknowledgments
\begin{sloppypar}
W. R. Assis is grateful to FAPESP (grant no. 2019/10239-7), and E. M. Franklin is grateful to FAPESP (grant no. 2018/14981-7) and to CNPq (grant no. 400284/2016-2) for the financial support provided. The authors would like to thank Fernando David C\'u\~nez for the assistance with the image processing code. Data supporting this work are available in the supporting information and in http://dx.doi.org/10.17632/f9p59sxm4f).
\end{sloppypar}

\bibliography{references}

\end{document}